\documentclass[12pt]{article}
\usepackage{amssymb,amsmath}
\usepackage[left=2cm,right=2cm,top=2cm,bottom=2cm]{geometry}
\usepackage{booktabs}
\usepackage{graphicx}
\usepackage{subcaption}
\usepackage{hyperref}
\usepackage{lscape}
\usepackage{enumitem}

\title{Bunch--excited wakefield in dielectric waveguide with hollow plasma channel}

\author{K.V. Galaydych\thanks{Corresponding author: kgalaydych@gmail.com},\,\,P.I. Markov,\,\,G.V. Sotnikov}

\date{National Science Center Kharkiv Institute of Physics and Technology 1, Akademichna St., Kharkiv, 61108, Ukraine}

\begin{document}

\maketitle

\section*{Abstract}
Wakefield excitation by a single relativistic electron bunch in a plasma--dielectric accelerating structure has been studied both analytically and numerically. The structure represents a dielectric--loaded cylindrical metal waveguide, which has partially plasma--filled channel (the hollow plasma channel) to transport charged particles. Assuming the linear regime of excitation, analytical expres\-sions have been derived for the longitudinal and radial wakefields generated by a finite--size electron bunch. Axial profiles of wakefield component amplitudes have been studied, and their mode and spectrum analyses have been performed. Furthermore, the electron bunch--driven wakefield excitation has been PIC--simulated numerically for the quasi--linear regime. The comparative analysis of the data resulting from analytical studies and the ones obtained by numerical simulation has demonstrated qualitative agreement between the results.

\section{Introduction}
The wakefield acceleration of charged particles in dielectric structures~\cite{Jing2016} ranges among new methods such as PWFA~\cite{Hogan2016,Adli2016}, LWFA~\cite{Nakajima2016} and DLA~\cite{Wootton2016,England2022}, and can be considered as one of the potential alternatives to the classical methods. Presently, all these methods continue to be currently central in the research, and are actively developed showing steady progress. The papers ~\cite{Jing2022,Cros2017,alegrocollaboration2019advanced,Albert2021,gourlay2022snowmass21,EuropeanStrategy,hidding2019plasma} presents road maps of future studies in these directions for  a wide range of problems and challenges in the physics of accelerators and charged particle beams. Thus, in dielectric structures, owing to the use of ceramic and polycrystalline materials (artificial diamond, quartz) that exhibit a high--level breakdown field, and also, due to advance in production of short (about a few tens of microns) electron bunches with a high charge value (up to a few tens of nC), it appears possible to attain high--amplitude excited fields. The experiments ~\cite{Thompson2008} have demonstrated the possibility of generating GV/m wakefields in the THz waveguide. In these structures, the wakefield is excited by relativistic high--charge drive bunches. In its passage through the structure, the drive bunch excites the Cherenkov radiation, which being reflected from the metal coating of the structure, forms the longitudinal field behind the drive bunch. In turn, this field can be used to accelerate the test bunch having a lower charge as opposed to that of the drive bunch. The delay time of the test bunch injection is matched so that in the passage through the charged particles transport channel (being generally vacuum), the test bunch could stay in the accelerating phase of the field as long as possible. It has been proposed in~\cite{Sotnikov2014} that the particle transport channel should be filled with cold plasma. As demonstrated in~\cite{Sotnikov2014}, that gives rise to the focusing properties of the wakefield excited in this plasma--dielectric waveguide due to the Langmuir wave excitation.
Previous analytical studies of wakefields in the cylindrical plasma--dielectric waveguide were carried out for the case, where homogeneous plasma fully filled the charged particle transport channel. With that, consideration has been given to two important cases, namely, i) when the drive bunch is injected strictly along the waveguide axis~\cite{Sotnikov2014}, and ii) when the drive bunch is injected with initial transverse offset~\cite{GALAYDYCH2022}. It has been shown in~\cite{GALAYDYCH2022} that for the cylindrical plasma--dielectric accelerating structure the presence of the initial transverse offset of the drive bunch does not lead to BBU as opposed to the dielectric--loaded structure without plasma filling~\cite{GaiKanareykin,Li,BaturinZholents}. Such plasma--dielectric accelerating structures cannot fully replace the classical focusing systems, yet, owing to the intrinsic focusing properties, they can substantially reduce the requirements to the external magnetic structures. By using PIC simulation in~\cite{Sotnikov2020,Markov2022}, consideration has been given to the non--linear regime of wakefield acceleration of test electron and positron bunches for the cases of homogeneous and inhomogeneous plasmas, where the electron bunch was considered to be the driver. However, despite the possibility of investigating the processes of excitation and acceleration through the use of the PIC simulation, a number of problems still remain to be solved. The need to have a clearer interpretation of the simulation results calls for more detailed analytical studies and development of the corresponding theory. Among the questions that remain to be answered we mention, for example, the question about the particular mechanism of formation of the focusing transverse field in the plasma region for the linear regime. The numerical analysis based on the analytical results makes it possible to carry out scanning and optimization of the parameters for the dielectric waveguide, plasma and drive bunch in an effort to determine the parameters that would provide steady test bunch acceleration. Besides, the feasible preliminary matching of the parameters appears to be a rather useful tool of theoretical analysis, as it can reduce the amount of numerical simulation. It is pertinent to note that the numerical simulation, being capable of taking into account many factors simultaneously, provides, on the one hand, a more comprehensive (when compared to the theory) description of the processes under study, but on the other hand, presents a computational extremely resource-intensive for the plasma systems comprised of a great number of particles. It is therefore only natural in the analytical studies to turn to the case of inhomogeneous plasma.
As the inhomogeneous plasma, for which the analytical theory can be built without using any approximation methods, we have chosen the case, where the transport channel is filled with homogeneous plasma incompletely, and along the waveguide axis there remains a cylindrical vacuum region. Compared to the waveguide with the fully plasma-filled transport channel, in the waveguide with incompletely filled channel one should expect the occurrence of additional eigenwaves, which, in turn, can be excited by the drive bunch, and thus can contribute to the excited wakefield.
This profile of plasma filling, known as the hollow plasma channel,  may take place, for example, at the capillary discharge~\cite{Ehrlich1996,Steinhauer2006}, or, when creating plasma by the use of a laser~\cite{Krushelnick1997} having the annular cross--section of its intensity profile. It should be noted that by themselves, the hollow plasma channels as wakefield acceleration structures show much promise and are of great interest because of the possibility to accelerate positrons~\cite{Schroeder1999,Lee2001,Gessner2016,Lindstrom2018}. For this reason, their intensive studies as promising acceleration structures are being continued, and the results obtained have been presented in numerous publications~\cite{GessnerPhD,LindstromPhD}.

The present paper has been focused on building a linear theory of wakefield excitation by a drive charged particle bunch in a cylindrical dielectric waveguide with a hollow plasma channel.

The paper is organized as follows. The statement of problem is formulated in Section~\ref{section:2}. Section~\ref{section:3} describes analytical investigations.  Section~\ref{section:4} deals with numerical analysis based on the results described in Section~\ref{section:3}. Section~\ref{section:5} gives the summary.

\section{Statement of problem}\label{section:2}
\begin{figure}[!th]
  \centering
  \includegraphics[width=0.5\textwidth]{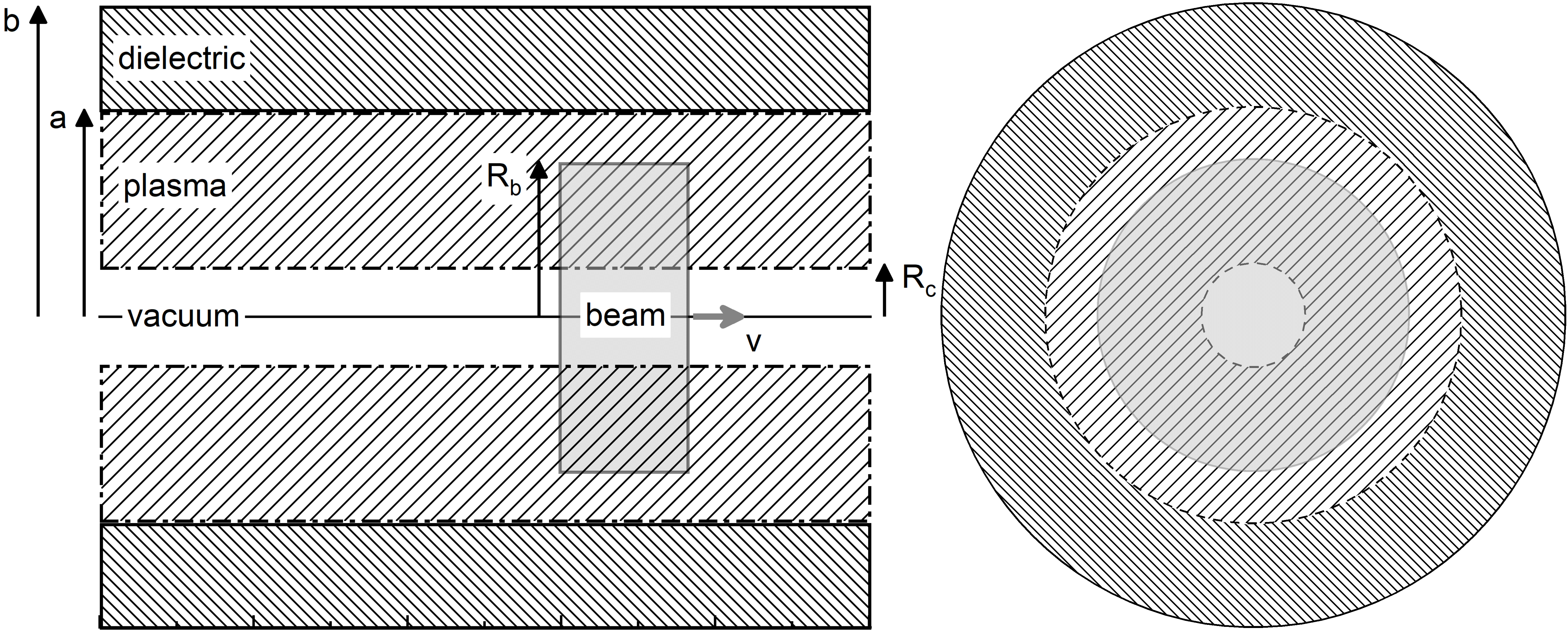}
  \caption{General view of the round plasma--dielectric waveguide. Metal coating, dielectric, plasma and drive beam are shown schematically. The on--axis drive beam moves along to the waveguide axis.}\label{Fig:01}
\end{figure}

The accelerating structure under consideration represents a cylindrical metal waveguide of radius $b$, partially filled with homogeneous isotropic dielectric having the relative permittivity $\varepsilon_d$. Inside the dielectric, there is a vacuum channel of radius $a$, partially filled with homogeneous isotropic plasma of density $n_p$. In turn, the plasma has a vacuum channel of radius $R_c$, and forms in this way a hollow plasma channel. Along the axis of the accelerating structure, at a constant velocity $v$ there propagates a non-evolving drive electron bunch of charge $Q_b$, length $L_b$ and radius $R_b$. In its passage the electron bunch excites the wakefield. In the general case, the drive bunch radius is larger than the vacuum channel radius. So, during its propagation, one part of the drive bunch is localized within the vacuum region, whereas the other is in the plasma region. Figure~\ref{Fig:01} schematically shows the waveguide, the dielectric, the hollow plasma channel and the drive bunch as longitudinal and transverse cross sections.
The plasma ions are assumed to be immobile, and the thermal motion of plasma electrons is neglected in paper. The vacuum-plasma boundary is taken to be sharp. It is also assumed that the dielectric has neither frequency nor spatial dispersion. The material of the metal conductor is assumed to be the ideal conductor. It is suggested that the density of the drive electron bunch is low in comparison with the plasma density, enabling us to consider the plasma as a linear medium. The external magnetic field is absent. The main goal of the present paper is to construct a linear analytical theory of wakefield excitation by a single electron bunch in the dielectric waveguide with a hollow plasma channel. The electrodynamics problem of structure excitation will be solved here in waveguide statement without taking into account the effects related to the longitudinal boundedness of the structure.

\section{Analytical studies}\label{section:3}
The theory of wakefield excitation by the drive electron bunch in the waveguide under study will be constructed in cylindrical coordinates $(r,\varphi,z)$ with the $z$--axis directed along the waveguide axis. According to the problem description: (I) the drive bunch propagates in the accelerating structure without any initial transverse offset, (II) only the longitudinal component of the drive bunch particle velocity vector is nonzero, (III) the charge distribution of the drive bunch is azimuthally homogeneous. Therefore, the nonzero contribution to the total wakefield will come only from the eigen $TM$--waves $(E_r,\,E_z,\,H_\varphi)$ of the bunch--excited waveguide under consideration.

The theory will be developed as follows. First, we determine the wakefield excited by the point--like particle (Green's function) of charge $q_0$ with the transverse coordinates $(r_0,\,\varphi_0)$. The particle is injected into the accelerating structure ($z=0$) at an arrival time $t_0$, and is moving at a constant velocity $v$. With the knowledge of the point--like particle--excited wakefield, we can derive the wakefield excited by the drive bunch of finite longitudinal and transverse size values by integrating over the injection time and transverse coordinate values of the particles.

Before proceeding to deriving the Green's function, it is of importance to note the following. The primary emphasis in the present study has been given to the excited electromagnetic fields, the structure of which is specified only by the external source, and no account has been taken of the effects that are associated with the longitudinal boundedness of the waveguide structure. When constructing the linear theory of excitation, we assume that the time and longitudinal coordinate dependence of the wakefield is determined in the same way as that for the source exciting the mentioned field.

The initial set of equations that describes the wakefield excitation represents the system of Maxwell equations for the radial component of the electric field $E_r$,\,the axial component of the electric field $E_z$,\,and the azimuthal component of the magnetic field $H_\varphi$
\begin{equation}\label{eq:01}
\begin{split}
&-\frac{1}{v}\frac{\partial E_r}{\partial \xi} - \frac{\partial E_z}{\partial r} = -\frac{1}{c}\frac{\partial H_\varphi}{\partial \xi},\\
&\frac{1}{v}\frac{\partial H_\varphi}{\partial \xi} = \frac{1}{c}\frac{\partial D_r}{\partial \xi},\\
&\frac{1}{r}\frac{\partial}{\partial r}\left( rH_\varphi \right) = \frac{4\pi}{c}j_z + \frac{1}{c}\frac{\partial D_z}{\partial \xi}
\end{split}
\end{equation}
where the densities of charge $\rho$ and of current $j_z$, created by the point--like charged particle are, respectively, equal to
\begin{equation}\label{eq:02}
\begin{split}
\rho &= \frac{q_0}{v}\frac{{\delta (r - {r_0})}}{r}\delta (\varphi  - {\varphi _0})\delta (\xi - {t_0}),\\
j_r &= 0,\,j_\varphi = 0,\,j_z = \rho v,
\end{split}
\end{equation}
where $c$ is the speed of light, and $\xi = t - z/v$. The formulated problem is uniform in both the time and direction of the drive particle motion. The excited wakefield, and also, the densities of charge and current can be written as the Fourier integrals over the variable $\xi$
\begin{equation}\label{eq:03}
\begin{split}
\begin{pmatrix}\mathbf{E}(r,\xi)\\\mathbf{D}(r,\xi)\\\mathbf{H}(r,\xi)\\\end{pmatrix}&= \int\limits_{ - \infty }^{ + \infty } {d\omega \begin{pmatrix}\mathbf{E}^\omega (r,\omega )\\\varepsilon(\omega)\mathbf{E}^\omega (r,\omega )\\\mathbf{H}^\omega (r,\omega )\\\end{pmatrix} }{e^{ - i\omega \xi}}{\text{,}}\\
\begin{pmatrix}\rho(r,\xi)\\j_z(r,\xi)\\\end{pmatrix}&= \int\limits_{ - \infty }^{ + \infty } {d\omega \begin{pmatrix}\rho^\omega (r,\omega )\\j_z^\omega (r,\omega )\\\end{pmatrix} }{e^{ - i\omega \xi}}{\text{.}}
\end{split}
\end{equation}
The dielectric permittivity in the vacuum regions ($0 \leq r \leq R_c$), plasma ($R_c \leq r \leq a$) and dielectric ($a \leq r \leq b$) is, respectively, equal to
\begin{equation}\label{eq:05}
\begin{split}
\varepsilon(\omega) =\begin{cases}
1,&\text{$0 \leq r \leq R_c$;}\\
\varepsilon _p(\omega ) = 1 - \omega _p^2/{\omega ^2},&\text{$R_c \leq r \leq a$;}\\
\varepsilon_d,&\text{$a \leq r \leq b$.}
\end{cases}
\end{split}
\end{equation}
where $\omega _p$ is the plasma frequency.

From equations~(\ref{eq:01}) we obtain the inhomogeneous wave equation for the Fourier transform of the axial component of the electric field $E_z^\omega$
\begin{equation}\label{eq:06}
\begin{split}
\frac{1}{r}\frac{\partial }{{\partial r}}\left( {r\frac{{\partial E_{z}^\omega }}{{\partial r}}} \right) - \frac{{{\omega ^2}}}{{{v^2}}}(1 - {\beta ^2}\varepsilon (\omega ))E_{z}^\omega  = \frac{{4\pi \omega i}}{{{v^2}}}\,\frac{{1 - {\beta ^2}\varepsilon (\omega )}}{{\varepsilon (\omega )}}j_{z}^\omega.
\end{split}
\end{equation}
In the right--hand side of ~(\ref{eq:06}) we have the Fourier transform of the axial component of the current density $j_z^\omega$, which can be written as
\begin{equation}\label{eq:07}
\begin{split}
j_{z}^\omega = \frac{q_0e^{i\omega t_0}}{4\pi^2}\int\limits_{0}^{+ \infty} J_0(\lambda r)J_0(\lambda r_0)\lambda d\lambda.
\end{split}
\end{equation}

In the case that the radius of drive bunch is larger than the radius of vacuum channel, the process of wakefield excitation is contributed by both the bunch particles being in the vacuum region and by the ones being in the plasma region. Below it will be shown that the frequency contents of the wakefield (and consequently, its spatiotemporal structure) is dependent on the excitation source location. Since the consideration is given here to the linear regime of excitation, then the total wakefield may be represented as a superposition of wakefields excited independently by the particles, being both in the vacuum and in the plasma.

The right--hand side of the wave equation for the Fourier transform of the axial component of the electric field $E_{z}^{\omega (I)}(r,\omega ,r_0,t_0)$, that is excited by the point--like charged particle propagating in vacuum is nonzero only in the vacuum region and is equal to $4\pi \omega i(1 - {\beta ^2})j_{z}^\omega/v^2$. In the case when the source of wakefield excitation is a charged particle propagating in the plasma, the right part of the equation for the Fourier transform of the axial component of the electric field $E_{z}^{\omega (II)}(r,\omega ,r_0,t_0)$ is nonzero only in the plasma region, and is equal to $4\pi \omega i(1 - {\beta ^2\varepsilon_p (\omega )})j_{z}^\omega/v^2\varepsilon_p (\omega )$.

Equation~(\ref{eq:06}) must be supplemented by the boundary conditions satisfied by the electromagnetic field excited in the waveguide. For the above--defined problem, these conditions are as follows. The amplitudes of wakefield components are finite at $r = 0$. On the surface of the perfectly conducting metal coating, the longitudinal electric field component turns to zero $E_{z}^{\omega (I,II)} (r=b)=0$. At vacuum--plasma and plasma--dielectric boundaries, the axial electric field component and the azimuthal magnetic field component are continuous $E_{z}^{\omega (I,II)} (r=R_c-0)=E_{z}^{\omega (I,II)} (r=R_c+0)$, $H_{\varphi}^{\omega (I,II)} (r=R_c-0)=H_{\varphi}^{\omega (I,II)} (r=R_c+0)$, $E_{z}^{\omega (I,II)} (r=a-0)=E_{z}^{\omega (I,II)} (r=a+0)$, $H_{\varphi}^{\omega (I,II)} (r=a-0)=H_{\varphi}^{\omega (I,II)} (r=a+0)$.

With the use of the Maxwell equations, the Fourier transforms of the radial electric field $E_{r}^\omega$, the axial magnetic field $H_{\varphi }^\omega$, and as a result, their combination $W_{r}^\omega = E_{r}^\omega - \beta H_{\varphi }^\omega$, which is involved in the radial force, can be expressed in terms of the Fourier transform of the axial electric field $E_{z}^\omega$
\begin{equation}\label{eq:10}
\begin{split}
&E_{r}^\omega  = - i\frac{v/\omega }{{(1 - {\beta ^2}\varepsilon (\omega ))}} {  \frac{{\partial E_{z}^\omega }}{{\partial r}}},\\
&H_{\varphi }^\omega  = - i\beta \varepsilon (\omega )\frac{v/\omega }{{(1 - {\beta ^2}\varepsilon (\omega ))}}{  \frac{{\partial E_{z}^\omega }}{{\partial r}}},\\
&W_{r}^\omega = E_{r}^\omega - \beta H_{\varphi }^\omega = -i\frac{v}{\omega}\frac{\partial E_{z}^\omega}{\partial r}
\end{split}
\end{equation}

Equation~(\ref{eq:06}) can be solved by the superposition a particular solution of the inhomogeneous equation and the general solution of the corresponding homogeneous equation
\begin{equation}\label{eq:11}
\begin{split}
E_{z}^{\omega (I)}(0 \leq r < r_0) = -i\frac{q_0\varkappa_{v}^2}{\pi\omega}\frac{I_0(\varkappa_{v}r)}{I_0(\varkappa_{v}R_c)}
\Delta_{0}(\varkappa_{v}R_c,\varkappa_{v}r_0)e^{i\omega t_0} -\\
i\frac{q_{0}}{\pi\omega R_c}\frac{I_0(\varkappa_{v}r_0)I_0(\varkappa_{v}r)}{I_0^2(\varkappa_{v}R_c)}\frac{D_2(\omega)}{D(\omega)}e^{i\omega t_0}
\end{split}
\end{equation}

\begin{equation}\label{eq:12}
\begin{split}
E_{z}^{\omega (I)}(r_0 < r \leq R_c) = -i\frac{q_0\varkappa_{v}^2}{\pi\omega}\frac{I_0(\varkappa_{v}r_0)}{I_0(\varkappa_{v}R_c)}\Delta_{0}(\varkappa_{v}R_c,\varkappa_{v}r)e^{i\omega t_0} -\\
i\frac{q_{0}}{\pi\omega R_c}\frac{I_0(\varkappa_{v}r_0)I_0(\varkappa_{v}r)}{I_0^2(\varkappa_{v}R_c)}\frac{D_2(\omega)}{D(\omega)}e^{i\omega t_0}
\end{split}
\end{equation}

\begin{equation}\label{eq:13}
\begin{split}
E_{z}^{\omega (I)}(R_c \leq r \leq a) = -i\frac{q_0}{\pi\omega R_c}\frac{I_0(\varkappa_{v}r_0)}{I_0(\varkappa_{v}R_c)}
\frac{D_2(\omega)}{D(\omega)}\frac{\Delta_{0}(\varkappa_{v}r,\varkappa_{v}a)}{\Delta_{0}(\varkappa_{v}R_c,\varkappa_{v}a)}e^{i\omega t_0} -\\
i\frac{q_0\varepsilon_p(\omega)}{\pi\omega \varkappa_{p}^2aR_cD(\omega)}\frac{I_0(\varkappa_{v}r_0)}{I_0(\varkappa_{v}R_c)}
\frac{\Delta_{0}(\varkappa_{p}r,\varkappa_{p}R_c)}{\Delta_{0}^2(\varkappa_{p}a,\varkappa_{p}R_c)}e^{i\omega t_0}
\end{split}
\end{equation}

\begin{equation}\label{eq:14}
\begin{split}
E_z^{\omega (II)}(0 \leq r \leq R_c) = \frac{ic}{\omega D(\omega)}\frac{I_0(\varkappa r)}{I_0(\varkappa R_c)}\left( f_1D_2-\frac{f_2\varepsilon_p}{\varkappa_p^2R_c\Delta_0(\varkappa_pa,\varkappa_pR_c)} \right)
\end{split}
\end{equation}

\begin{equation}\label{eq:15}
\begin{split}
E_z^{\omega (II)}(R_c \leq r \leq r_0) = \frac{iq_0\varkappa_p^2}{\pi\omega\varepsilon_p}G_p(r,r_0,\omega)e^{i\omega t_0} +\\
\frac{ic}{\omega D(\omega)}\left( f_1D_2 - \frac{f_2\varepsilon_p}{\varkappa_p^2R_c\Delta_0(\varkappa_pa,\varkappa_pR_c)} \right)\frac{\Delta_0(\varkappa_pr,\varkappa_pa)}{\Delta_0(\varkappa_pR_c,\varkappa_pa)} -\\
\frac{ic}{\omega D(\omega)}\left( f_2D_1 + \frac{f_1\varepsilon_p}{\varkappa_p^2a\Delta_0(\varkappa_pR_c,\varkappa_pa)} \right)\frac{\Delta_0(\varkappa_pr,\varkappa_pR_c)}{\Delta_0(\varkappa_pa,\varkappa_pR_c)}
\end{split}
\end{equation}

\begin{equation}\label{eq:16}
\begin{split}
E_z^{\omega (II)}(r_0 \leq r \leq a) = \frac{iq_0\varkappa_p^2}{\pi\omega\varepsilon_p}H_p(r,r_0,\omega)e^{i\omega t_0} +\\
\frac{ic}{\omega D(\omega)}\left( f_1D_2 - \frac{f_2\varepsilon_p}{\varkappa_p^2R_c\Delta_0(\varkappa_pa,\varkappa_pR_c)} \right)\frac{\Delta_0(\varkappa_pr,\varkappa_pa)}{\Delta_0(\varkappa_pR_c,\varkappa_pa)} -\\
\frac{ic}{\omega D(\omega)}\left( f_2D_1 + \frac{f_1\varepsilon_p}{\varkappa_p^2a\Delta_0(\varkappa_pR_c,\varkappa_pa)} \right)\frac{\Delta_0(\varkappa_pr,\varkappa_pR_c)}{\Delta_0(\varkappa_pa,\varkappa_pR_c)},
\end{split}
\end{equation}
In expressions~(\ref{eq:11})--~(\ref{eq:16}) the following designations are used:
\begin{equation}
\begin{split}
\varkappa _{v}^2 = (\omega/v)^2(1 - {\beta ^2}),\,\varkappa _{p}^2 = (\omega/v)^2(1 - {\beta ^2}{\varepsilon _p}({\omega})),
\end{split}
\end{equation}
\begin{equation}
\begin{split}
\Delta _0(x,y) = I_0(x)K_0(y) - K_0(x)I_0(y),\,\Delta _1(x,y) = I_1(x)K_0(y) + K_1(x)I_0(y),
\end{split}
\end{equation}
where $I_{0,1}$ and $K_{0,1}$ are the modified Bessel and MacDonald functions of the zero and first order, and the functions $f_1(\omega)$, $f_2(\omega)$, $G_p(r,r_0,\omega)$, $H_p(r,r_0,\omega)$, $D(\omega)$ take the following forms:
\begin{equation}
\begin{split}
f_1(\omega) = -\frac{q_0e^{i\omega t_0}}{\pi cR_c}\frac{\Delta_{0}(\varkappa_{p}r_0,\varkappa_{p}a)}{\Delta_{0}(\varkappa_{ps}R_c,\varkappa_{ps}a)},\,f_2(\omega) = \frac{q_0e^{i\omega t_0}}{\pi ca}\frac{\Delta_{0}(\varkappa_{p}r_0,\varkappa_{p}R_c)}{\Delta_{0}(\varkappa_{ps}a,\varkappa_{ps}R_c)},
\end{split}
\end{equation}

\begin{equation}
\begin{split}
G_p(r,r_0,\omega) = \frac{\Delta_{0}(\varkappa_{p}r,\varkappa_{p}R_c)\Delta_{0}(\varkappa_{p}a,\varkappa_{p}r_0)}{\Delta_{0}(\varkappa_{p}R_c,\varkappa_{p}a)},\,H_p(r,r_0,\omega) = \frac{\Delta_{0}(\varkappa_{p}r,\varkappa_{p}a)\Delta_{0}(\varkappa_{p}r_0,\varkappa_{p}R_c)}{\Delta_{0}(\varkappa_{p}a,\varkappa_{p}R_c)},
\end{split}
\end{equation}

\begin{equation}\label{eq:18}
\begin{split}
D(\omega) &= D_1(\omega)D_2(\omega)+D_3(\omega),\\
D_1(\omega) &= \frac{1}{\varkappa_{v}}\frac{I_1(\varkappa_{v}R_c)}{I_0(\varkappa_{v}R_c)} - \frac{\varepsilon_p}{\varkappa_{p}}\frac{\Delta_1(\varkappa_{p}R_c,\varkappa_{p}a)}{\Delta_0(\varkappa_{p}R_c,\varkappa_{p}a)},\\
D_2(\omega) &= \frac{\varepsilon_p}{\varkappa_{p}}\frac{\Delta_1(\varkappa_{p}a,\varkappa_{p}R_c)}{\Delta_0(\varkappa_{p}a,\varkappa_{p}R_c)} + \frac{\varepsilon_d}{\varkappa_{d}}\frac{F_1(\varkappa_{d}a,\varkappa_{d}b)}{F_0(\varkappa_{d}a,\varkappa_{d}b)},\\
D_3(\omega) &= \frac{\varepsilon_p^2}{\varkappa_{p}^2}\frac{\Delta_1(\varkappa_{p}R_c,\varkappa_{p}R_c)}{\Delta_0(\varkappa_{p}a,\varkappa_{p}R_c)}
\frac{\Delta_1(\varkappa_{p}a,\varkappa_{p}a)}{\Delta_0(\varkappa_{p}R_c,\varkappa_{p}a)},
\end{split}
\end{equation}
and $\varkappa_d^2=(\omega/v)^2(\beta^2\varepsilon_d-1)$.

By integrating over the frequency~(\ref{eq:03}) we obtain the expressions for the Green's functions of the longitudinal component of the electric field excited by point--like charged particles, which are propagating in the vacuum channel $E_{Gz}^{(I)}(r,\xi,r_0,t_0)$, and in the plasma $E_{Gz}^{(II)}(r,\xi,r_0,t_0)$
\begin{equation}\label{eq:19}
\begin{split}
E_{Gz}^{(I)}(0 \leq r \leq R_c) = -\sum\limits_{s = 1}^{+ \infty}\frac{4q_0\varkappa_{vs}}{\omega_sR_c}\frac{I_0(\varkappa_{vs}r)I_0(\varkappa_{vs}r_0)}{I_0^2(\varkappa_{vs}R_c)}
\frac{D_{2s}}{D'({\omega_s})}\times\\
cos\omega_s(\xi-t_0)\theta(\xi-t_0)
\end{split}
\end{equation}

\begin{equation}\label{eq:20}
\begin{split}
E_{Gz}^{(I)}(R_c \leq r \leq a) = -\sum\limits_{s = 1}^{+ \infty}\frac{4q_0}{\omega_sD'({\omega _s})R_c}\frac{I_0(\varkappa_{vs}r_0)}{I_0(\varkappa_{vs}R_c)}cos\omega_s(\xi-t_0)\theta(\xi-t_0)\times\\
\left(D_{2s}\frac{\Delta_{0}(\varkappa_{ps}r,\varkappa_{ps}a)}{\Delta_{0}(\varkappa_{ps}R_c,\varkappa_{ps}a)}-
\frac{\varepsilon_{ps}}{\varkappa_{ps}^2a}\frac{\Delta_{0}(\varkappa_{ps}r,\varkappa_{ps}R_c)}
{\Delta_{0}(\varkappa_{ps}a,\varkappa_{ps}R_c)}\right)
\end{split}
\end{equation}

\begin{equation}\label{eq:21}
\begin{split}
W_{Gr}^{(I)}(0 \leq r \leq R_c) = \sum\limits_{s = 1}^{+ \infty}\frac{4q_{0}v\varkappa_{vs}}{\omega_{s}^2R_c}\frac{I_1(\varkappa_{vs}r)I_0(\varkappa_{vs}r_0)}{I_0^2(\varkappa_{vs}R_c)}\times\\
\frac{D_{2s}}{D'({\omega_s})}sin\omega_s(\xi-t_0)\theta(\xi-t_0)
\end{split}
\end{equation}

\begin{equation}\label{eq:22}
\begin{split}
W_{Gr}^{(I)}(R_c \leq r \leq a) = \sum\limits_{s = 1}^{+ \infty}\frac{4q_{0}v\varkappa_{ps}}{\omega_{s}^2D'(\omega_s)}\frac{I_0(\varkappa_{vs}r_0)}{I_0(\varkappa_{vs}R_c)}\times\\
\left( \frac{D_{2s}}{R_c}\frac{\Delta_{1}(\varkappa_{ps}r,\varkappa_{ps}a)}{\Delta_{0}(\varkappa_{ps}R_c,\varkappa_{ps}a)} - \frac{\varepsilon_{ps}}{\varkappa_{ps}^2aR_c\Delta_{0}(\varkappa_{ps}R_c,\varkappa_{ps}a)}
\frac{\Delta_{1}(\varkappa_{ps}r,\varkappa_{ps}R_c)}{\Delta_{0}(\varkappa_{ps}a,\varkappa_{ps}R_c)} \right)\times\\
sin\omega_s(\xi-t_0)\theta(\xi-t_0)
\end{split}
\end{equation}

\begin{equation}\label{eq:23}
\begin{split}
E_{Gz}^{(II)}(0 \leq r \leq R_c) = -\sum\limits_{s = 1}^{+ \infty}\frac{4q_0}{\omega_sD'({\omega _s})}\frac{I_0(\varkappa_{vs}r)}{I_0(\varkappa_{vs}R_c)}cos\omega_s(\xi-t_0)\theta(\xi-t_0)\times\\
\left(\frac{D_{2s}\Delta_{0}(\varkappa_{ps}r_0,\varkappa_{ps}a)}{R_c\Delta_{0}(\varkappa_{ps}R_c,\varkappa_{ps}a)}
-\frac{\varkappa_{ps}^2D_{3s}}{\varepsilon_{ps}}\Delta_{0}(\varkappa_{ps}r_0,\varkappa_{ps}R_c)\right)
\end{split}
\end{equation}

\begin{equation}\label{eq:24}
\begin{split}
E_{Gz}^{(II)}(R_c \leq r \leq r_0) = 2q_0k_p^2G_p(r,r_0,\omega_p)cos\omega_p(\xi-t_0)\theta(\xi-t_0)-\\
\sum\limits_{s = 1}^{+\infty}\frac{4q_0}{\omega_sD'({\omega_s})}
\frac{\Delta_{0}(\varkappa_{ps}r,\varkappa_{ps}a)}{\Delta_{0}(\varkappa_{ps}R_c,\varkappa_{ps}a)}
cos\omega_s(\xi-t_0)\theta(\xi-t_0)\times\\
\left( \frac{D_{2s}\Delta_{0}(\varkappa_{ps}r_0,\varkappa_{ps}a)}{R_c\Delta_{0}(\varkappa_{ps}R_c,\varkappa_{ps}a)}-
\frac{\varkappa_{ps}^2D_{3s}}{\varepsilon_{ps}}\Delta_{0}(\varkappa_{ps}r_0,\varkappa_{ps}R_c) \right) -\\
\sum\limits_{s = 1}^{+ \infty}\frac{4q_0}{\omega_sD'(\omega _s)}\frac{\Delta_{0}(\varkappa_{ps}r,\varkappa_{ps}R_c)}{\Delta_{0}(\varkappa_{ps}a,\varkappa_{ps}R_c)}
cos\omega_s(\xi-t_0)\theta(\xi-t_0)\times\\
\left( \frac{D_{1s}\Delta_{0}(\varkappa_{ps}r_0,\varkappa_{ps}R_c)}{a\Delta_{0}(\varkappa_{ps}a,\varkappa_{ps}R_c)} + \frac{\varkappa_{ps}^2D_{3s}}{\varepsilon_{ps}}\Delta_{0}(\varkappa_{ps}r_0,\varkappa_{ps}a) \right)
\end{split}
\end{equation}

\begin{equation}\label{eq:25}
\begin{split}
E_{Gz}^{(II)}(r_0 \leq r \leq a) = 2q_0k_p^2H_p(r,r_0,\omega_p)cos\omega_p(\xi-t_0)\theta(\xi-t_0)-\\
\sum\limits_{s = 1}^{+\infty}\frac{4q_0}{\omega_sD'({\omega_s})}
\frac{\Delta_{0}(\varkappa_{ps}r,\varkappa_{ps}a)}{\Delta_{0}(\varkappa_{ps}R_c,\varkappa_{ps}a)}
cos\omega_s(\xi-t_0)\theta(\xi-t_0)\times\\
\left( \frac{D_{2s}\Delta_{0}(\varkappa_{ps}r_0,\varkappa_{ps}a)}{R_c\Delta_{0}(\varkappa_{ps}R_c,\varkappa_{ps}a)}-
\frac{\varkappa_{ps}^2D_{3s}}{\varepsilon_{ps}}\Delta_{0}(\varkappa_{ps}r_0,\varkappa_{ps}R_c) \right) -\\
\sum\limits_{s = 1}^{+ \infty}\frac{4q_0}{\omega_sD'(\omega _s)}\frac{\Delta_{0}(\varkappa_{ps}r,\varkappa_{ps}R_c)}{\Delta_{0}(\varkappa_{ps}a,\varkappa_{ps}R_c)}
cos\omega_s(\xi-t_0)\theta(\xi-t_0)\times\\
\left( \frac{D_{1s}\Delta_{0}(\varkappa_{ps}r_0,\varkappa_{ps}R_c)}{a\Delta_{0}(\varkappa_{ps}a,\varkappa_{ps}R_c)} + \frac{\varkappa_{ps}^2D_{3s}}{\varepsilon_{ps}}\Delta_{0}(\varkappa_{ps}r_0,\varkappa_{ps}a) \right)
\end{split}
\end{equation}

\begin{equation}\label{eq:26}
\begin{split}
W_{Gr}^{(II)}(0 \leq r \leq R_c) = -\sum\limits_{s = 1}^{+ \infty}\frac{4\pi cv\varkappa_{vs}}{\omega_{s}^2D'(\omega_s)}\frac{I_1(\varkappa_{vs}r)}{I_0(\varkappa_{vs}R_c)}\times\\
\left( f_{1s}D_{2s} - \frac{f_{2s}\varepsilon_{ps}}{\varkappa_{ps}^2R_c\Delta_{0}(\varkappa_{p}a,\varkappa_{p}R_c)} \right)sin\omega_s(\xi-t_0)\theta(\xi-t_0)
\end{split}
\end{equation}

\begin{equation}\label{eq:27}
\begin{split}
W_{Gr}^{(II)}(R_c \leq r \leq r_0) = -2q_{0}k_p^2G'_p(\omega_p)\times\\
sin\omega_p(\xi-t_0)\theta(\xi-t_0)-\sum\limits_{s = 1}^{+ \infty}\frac{4\pi cv\varkappa_{ps}}{\omega_{s}^2D'(\omega_s)}
sin\omega_s(\xi-t_0)\theta(\xi-t_0)\times\\
\left( \frac{\Delta_{1}(\varkappa_{ps}r,\varkappa_{ps}a)}{\Delta_{0}(\varkappa_{ps}R_c,\varkappa_{ps}a)}\left\{ f_{1s}D_{2s} -
\frac{f_{2s}\varepsilon_{ps}}{\varkappa_{ps}^2R_c\Delta_{0}(\varkappa_{p}a,\varkappa_{p}R_c)} \right\} - \right. \\ \left. \frac{\Delta_{1}(\varkappa_{ps}r,\varkappa_{ps}R_c)}{\Delta_{0}(\varkappa_{ps}a,\varkappa_{ps}R_c)}\left\{ f_{2s}D_{1s} +
\frac{f_{1s}\varepsilon_{ps}}{\varkappa_{ps}^2a\Delta_{0}(\varkappa_{ps}R_c,\varkappa_{ps}a)} \right\} \right)
\end{split}
\end{equation}

\begin{equation}\label{eq:28}
\begin{split}
W_{Gr}^{(II)}(r_0 \leq r \leq a) = -2q_{0}k_p^2H'_p(\omega_p)\times\\
sin\omega_p(\xi-t_0)\theta(\xi-t_0)-\sum\limits_{s = 1}^{+ \infty}\frac{4\pi cv\varkappa_{ps}}{\omega_{s}^2D'(\omega_s)}
sin\omega_s(\xi-t_0)\theta(\xi-t_0)\times\\
\left( \frac{\Delta_{1}(\varkappa_{ps}r,\varkappa_{ps}a)}{\Delta_{0}(\varkappa_{ps}R_c,\varkappa_{ps}a)}\left\{ f_{1s}D_{2s} -
\frac{f_{2s}\varepsilon_{ps}}{\varkappa_{ps}^2R_c\Delta_{0}(\varkappa_{p}a,\varkappa_{p}R_c)} \right\} - \right. \\ \left. \frac{\Delta_{1}(\varkappa_{ps}r,\varkappa_{ps}R_c)}{\Delta_{0}(\varkappa_{ps}a,\varkappa_{ps}R_c)}\left\{ f_{2s}D_{1s} +
\frac{f_{1s}\varepsilon_{ps}}{\varkappa_{ps}^2a\Delta_{0}(\varkappa_{ps}R_c,\varkappa_{ps}a)} \right\} \right),
\end{split}
\end{equation}
where the index $s$ denotes that the corresponding quantities were calculated at $\omega = \omega_s$, while $D'(\omega_s)=\left( \frac{d D(\omega)}{d \omega} \right)_{\omega=\omega_s}.$

Expressions~(\ref{eq:19})--~(\ref{eq:28}) demonstrate the fact that the frequency content of the excited wakefield depends on the region of the waveguide in which the charged particle, which is its source, propagates. Namely, if the exciting particle is in the vacuum channel, then the expressions $E_{Gz}^{(I)}(r,\xi,r_0,t_0)$ for the regions of both the vacuum and the plasma, and $E_{Gz}^{(II)}(r,\xi,r_0,t_0)$ for the vacuum region, contain only the frequencies $\omega_s$, which are defined by the pole $D(\omega_s ) = 0$ in the integrands. The frequencies $\omega_s$ are the fundamental frequencies of the waveguide under study, which are in resonance with the bunch (Cherenkov synchronism). This exhibits the fact that on moving in vacuum, the charged particle does not excite the plasma Langmuir wave. Furthermore, the plasma Langmuir wave, excited at the frequency $\omega_p$ by the charged particle propagating in the plasma, is localized in the plasma. And in this case, the structure of the total wakefield is determined by the wakefields excited at the frequencies $\omega_s$ being synchronous with the drive bunch. While the expressions $E_{Gz}^{(II)}(r,\xi,r_0,t_0)$ in the plasma region are determined by both the resonant frequencies $\omega_s$, and the plasma frequency $\omega_p$, which is defined by the pole $\varepsilon_p (\omega_p ) = 0$. And in this case, the total wakefield is formed by both the wakefields at the frequencies $\omega_s$, being in synchronism with the bunch, and the wakefield of the plasma wave at the frequency $\omega_p$.

By integration over the times of arrival $t_0$, and also over the transverse coordinates of the particles $(r_0,\varphi_0)$ we obtain the final expressions for the longitudinal and transverse components of the wakefield excited by the drive bunch, that has the radius larger than the vacuum channel radius ($R_b>R_c$):

vacuum channel region $(0 \leq r \leq R_c)$
\begin{equation}\label{eq:29}
\begin{split}
E_z(r,\xi) = -\sum\limits_{s = 1}^{+ \infty}\frac{8Q_b}{\omega_sD'(\omega_s)R_b^2}
\frac{I_0(\varkappa_{vs}r)}{I_0(\varkappa_{vs}R_c)}\Psi_{\parallel}(\xi,\omega_s)\times\\
\left( \frac{D_{2s}I_1(\varkappa_{vs}R_c)}{\varkappa_{vs}I_0(\varkappa_{vs}R_c)} + \frac{D_{2s}\Phi_1(\varkappa_{ps}R_c)}{\varkappa_{ps}^2R_c\Delta_{0}(\varkappa_{ps}R_c,\varkappa_{ps}a)}-
\frac{D_{3s}\Phi_2(\varkappa_{ps}R_b)}{\varepsilon_{ps}} \right),
\end{split}
\end{equation}

\begin{equation}\label{eq:30}
\begin{split}
W_r(r,\xi) = \sum\limits_{s = 1}^{+ \infty}\frac{8Q_{b}v}{\omega_{s}^2D'(\omega_s)R_b^2}
\frac{I_1(\varkappa_{vs}r)}{I_0(\varkappa_{vs}R_c)}\Psi_{\perp}(\xi,\omega_s)\times\\
\left( D_{2s}\frac{I_1(\varkappa_{vs}R_c)}{I_0(\varkappa_{vs}R_c)} + \frac{\varkappa_{vs}}{\varkappa_{ps}^2R_c}\left\{ \frac{D_{2s}\Phi_{1}(\varkappa_{ps}R_c)}{\Delta_{0}(\varkappa_{ps}R_c,\varkappa_{ps}a)} +
\frac{\varepsilon_{ps}\Phi_{2}(\varkappa_{ps}R_b)}{\varkappa_{ps}^2a\Delta_{0}^2(\varkappa_{ps}a,\varkappa_{ps}R_c)} \right\} \right)
\end{split}
\end{equation}
the plasma region, which has the drive bunch in $(R_c \leq r \leq R_b)$
\begin{equation}\label{eq:31}
\begin{split}
E_z(r,\xi) = \frac{4Q_b}{R_b^2}\Psi_{\parallel}(\xi,\omega_p)\left( \frac{\Delta_{0}(k_{p}r,k_{p}a)}{\Delta_{0}(k_{p}a,k_{p}R_c)}\Phi_2(k_{p}r) + \frac{\Delta_{0}(k_{p}r,k_{p}R_c)}{\Delta_{0}(k_{p}a,k_{p}R_c)}\Phi_1(k_{p}r) \right)\\
 -\sum\limits_{s = 1}^{+ \infty}\frac{8Q_b}{\omega_sD'(\omega _s)R_b^2}\Psi_{\parallel}(\xi,\omega_s)\times\\
\left( \frac{I_1(\varkappa_{vs}R_b)}{\varkappa_{vs}I_0(\varkappa_{vs}R_c)}\left\{ D_{2s}\frac{\Delta_{0}(\varkappa_{ps}r,\varkappa_{ps}a)}{\Delta_{0}(\varkappa_{ps}R_c,\varkappa_{ps}a)} -
\frac{\varepsilon_{ps}}{\varkappa_{ps}^2a}\frac{\Delta_{0}(\varkappa_{ps}r,\varkappa_{ps}R_c)}
{\Delta_{0}(\varkappa_{ps}a,\varkappa_{ps}R_c)} \right\} + \right. \\ \left.
\frac{\Delta_{0}(\varkappa_{ps}r,\varkappa_{ps}a)}{\Delta_{0}(\varkappa_{ps}R_c,\varkappa_{ps}a)}\left\{ \frac{D_{2s}\Phi_1(\varkappa_{ps}R_c)}{\varkappa_{ps}^2R_c\Delta_{0}(\varkappa_{ps}R_c,\varkappa_{ps}a)} -
\frac{D_{3s}}{\varepsilon_{ps}}\Phi_2(\varkappa_{ps}R_b) \right\} + \right. \\ \left.
\frac{\Delta_{0}(\varkappa_{ps}r,\varkappa_{ps}R_c)}{\Delta_{0}(\varkappa_{ps}a,\varkappa_{ps}R_c)}\left\{ \frac{D_{1s}\Phi_2(\varkappa_{ps}R_b)}{\varkappa_{ps}^2a\Delta_{0}(\varkappa_{ps}a,\varkappa_{ps}R_c)} +
\frac{D_{3s}}{\varepsilon_{ps}}\Phi_1(\varkappa_{ps}R_c) \right\} \right)
\end{split}
\end{equation}

\begin{equation}\label{eq:32}
\begin{split}
W_r(r,\xi) = -\frac{4Q_b}{R_b^2}\Psi_{\perp}(\xi,\omega_p)\left( \frac{\Delta_{1}(k_{p}r,k_{p}a)}{\Delta_{0}(k_{p}a,k_{p}R_c)}\Phi_2(k_{p}r) + \frac{\Delta_{1}(k_{p}r,k_{p}R_c)}{\Delta_{0}(k_{p}a,k_{p}R_c)}\Phi_1(k_{p}r) \right) +\\
\sum\limits_{s = 1}^{+ \infty}\frac{8Q_{b}v}{\omega_{s}^2D'(\omega_s)R_b^2}
\Psi_{\perp}(\xi,\omega_s)\times\\
\left( \frac{\varkappa_{ps}I_1(\varkappa_{vs}R_c)}{\varkappa_{vs}I_0(\varkappa_{vs}R_c)}\left\{ D_{2s}\frac{\Delta_{1}(\varkappa_{ps}r,\varkappa_{ps}a)}{\Delta_{0}(\varkappa_{ps}R_c,\varkappa_{ps}a)}-
\frac{\varepsilon_{ps}}{\varkappa_{ps}^2a\Delta_{0}(\varkappa_{ps}R_c,\varkappa_{ps}a)}\frac{\Delta_{1}(\varkappa_{ps}r,\varkappa_{ps}R_c)}
{\Delta_{0}(\varkappa_{ps}a,\varkappa_{ps}R_c)} \right\} + \right. \\ \left.
\frac{\Delta_{1}(\varkappa_{ps}r,\varkappa_{ps}a)}{\Delta_{0}(\varkappa_{ps}R_c,\varkappa_{ps}a)}\left\{ \frac{D_{2s}\Phi_1(\varkappa_{ps}R_c)}{\varkappa_{ps}R_c\Delta_{0}(\varkappa_{ps}R_c,\varkappa_{ps}a)} +
\frac{\varepsilon_{ps}\Phi_2(\varkappa_{ps}R_b)}{\varkappa_{ps}^3aR_c\Delta_{0}^2(\varkappa_{ps}a,\varkappa_{ps}R_c)} \right\} + \right. \\ \left.
\frac{\Delta_{1}(\varkappa_{ps}r,\varkappa_{ps}R_c)}{\Delta_{0}(\varkappa_{ps}a,\varkappa_{ps}R_c)}\left\{ \frac{D_{1s}\Phi_2(\varkappa_{ps}R_b)}{\varkappa_{ps}a\Delta_{0}(\varkappa_{ps}a,\varkappa_{ps}R_c)} -
\frac{\varepsilon_{ps}\Phi_1(\varkappa_{ps}R_c)}{\varkappa_{ps}^3aR_c\Delta_{0}^2(\varkappa_{ps}R_c,\varkappa_{ps}a)} \right\} \right),
\end{split}
\end{equation}
where
\begin{equation}\label{eq:33}
\begin{split}
\Phi_1(\alpha r) &= \alpha R_b\Delta_{1}(\alpha R_b,\alpha a) - \alpha r\Delta_{1}(\alpha r,\alpha a),\\
\Phi_2(\alpha r) &= \alpha r\Delta_{1}(\alpha r,\alpha R_c) - \alpha R_c\Delta_{1}(\alpha R_c,\alpha R_c)
\end{split}
\end{equation}
The functions $\Psi_{\parallel}(\xi,\omega)$ and $\Psi_{\perp}(\xi,\omega)$, that describe the longitudinal structure of the axial and transverse  wakefield components take the following forms:
\begin{equation}\label{eq:34}
\begin{split}
\Psi_{\parallel}(\xi,\omega) = \frac{1}{\omega T_b}\bigg( \theta(\xi)sin\omega\xi - \theta(\xi - T_b)sin\omega(\xi - T_b) \bigg) -\\
\frac{1}{4\pi^2 - \omega^2 T_b^2}\left( \theta(\xi)\left\{2\pi sin\frac{2\pi}{T_b}\xi - \omega T_bsin\omega\xi\right\} -\right. \\ \left. \theta(\xi-T_b)\left\{2\pi sin\frac{2\pi}{T_b}(\xi-T_b) - \omega T_bsin\omega(\xi-T_b)\right\} \right),\\
\Psi_{\perp}(\xi,\omega) = \frac{1}{\omega T_b}\bigg( \theta(\xi)\Big(1-cos\omega\xi\Big) - \theta(\xi - T_b)\Big(1-cos\omega(\xi - T_b)\Big) \bigg) -\\
\frac{\omega T_b}{\omega^2 T_b^2 - 4\pi^2}\left( \theta(\xi)\left\{cos\frac{2\pi}{T_b}\xi - cos\omega\xi\right\} -\right. \\ \left. \theta(\xi-T_b)\left\{cos\frac{2\pi}{T_b}(\xi-T_b) - cos\omega(\xi-T_b)\right\} \right),
\end{split}
\end{equation}
where $T_{b} = L_{b}/v$ is the temporal length of the drive bunch.

\section{Numerical analysis}\label{section:4}
The analytical expressions derived for the longitudinal and transverse components of the azimuthally homogeneous wakefield, excited by a single relativistic drive electron bunch, make it possible to analyze the amplitude profiles distribution of these components, as well as to perform their mode and spectrum analyses. For numerical analysis, we have chosen the following parameters: inner dielectric radius $a = 500\,\mu m$, outer dielectric radius $b = 600\,\mu m$, vacuum channel radius $R_c = 200\,\mu m$, permittivity of dielectric $\varepsilon_d = 3.75$ (quartz), plasma density $n_p = 2.0\cdot 10^{14}\,cm^{-3}$ (plasma frequency $f_p=127.3\,GHz$), energy of electron drive bunch $5\,GeV$, charge of the drive bunch $Q_b = 3.0\, nC$, length of the drive bunch $L_b = 250\,\mu m$, radius of the drive bunch $R_b = 450\,\mu m$.

\begin{figure}[!th]
  \centering
  \includegraphics[width=0.49\textwidth]{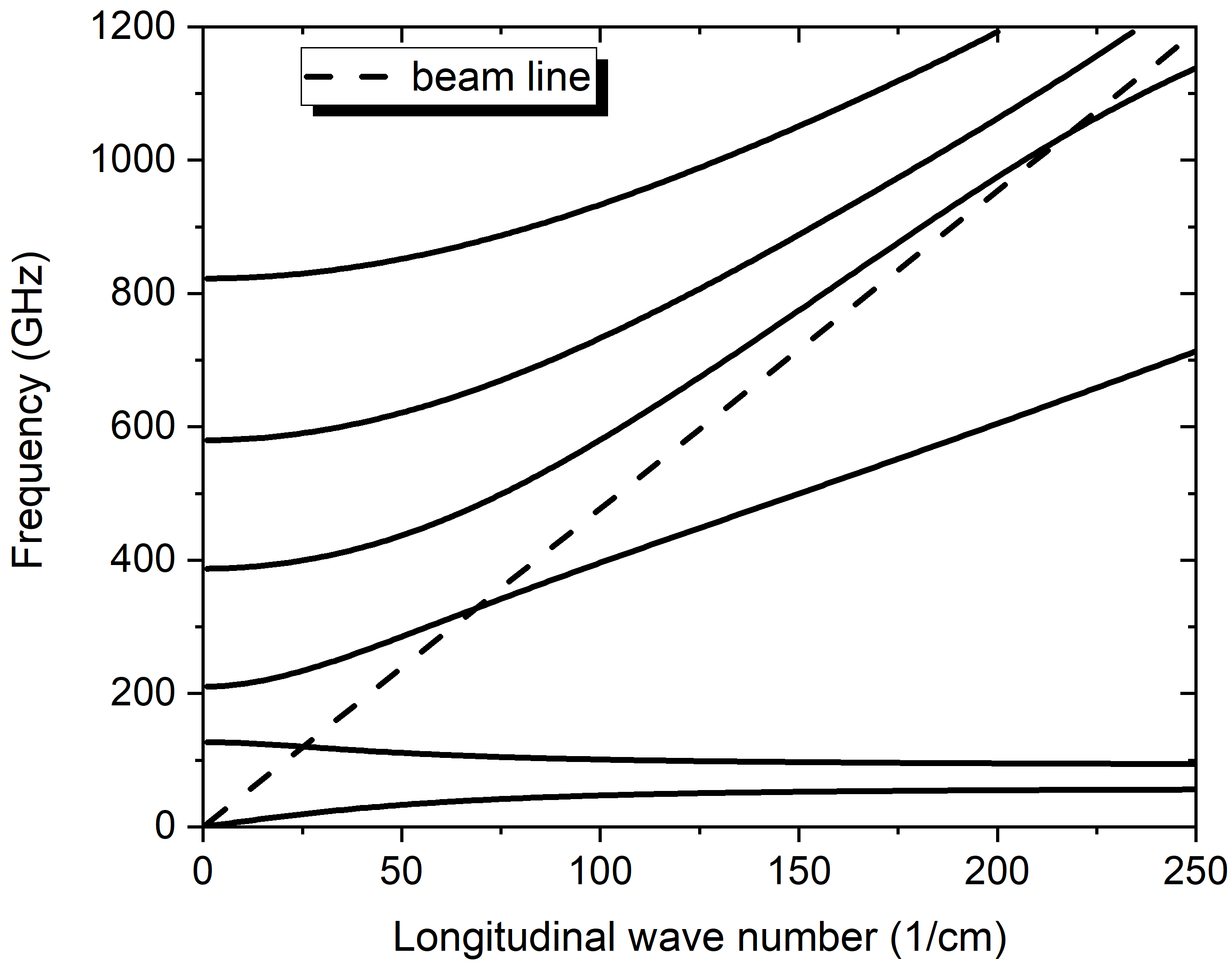}
  \includegraphics[width=0.49\textwidth]{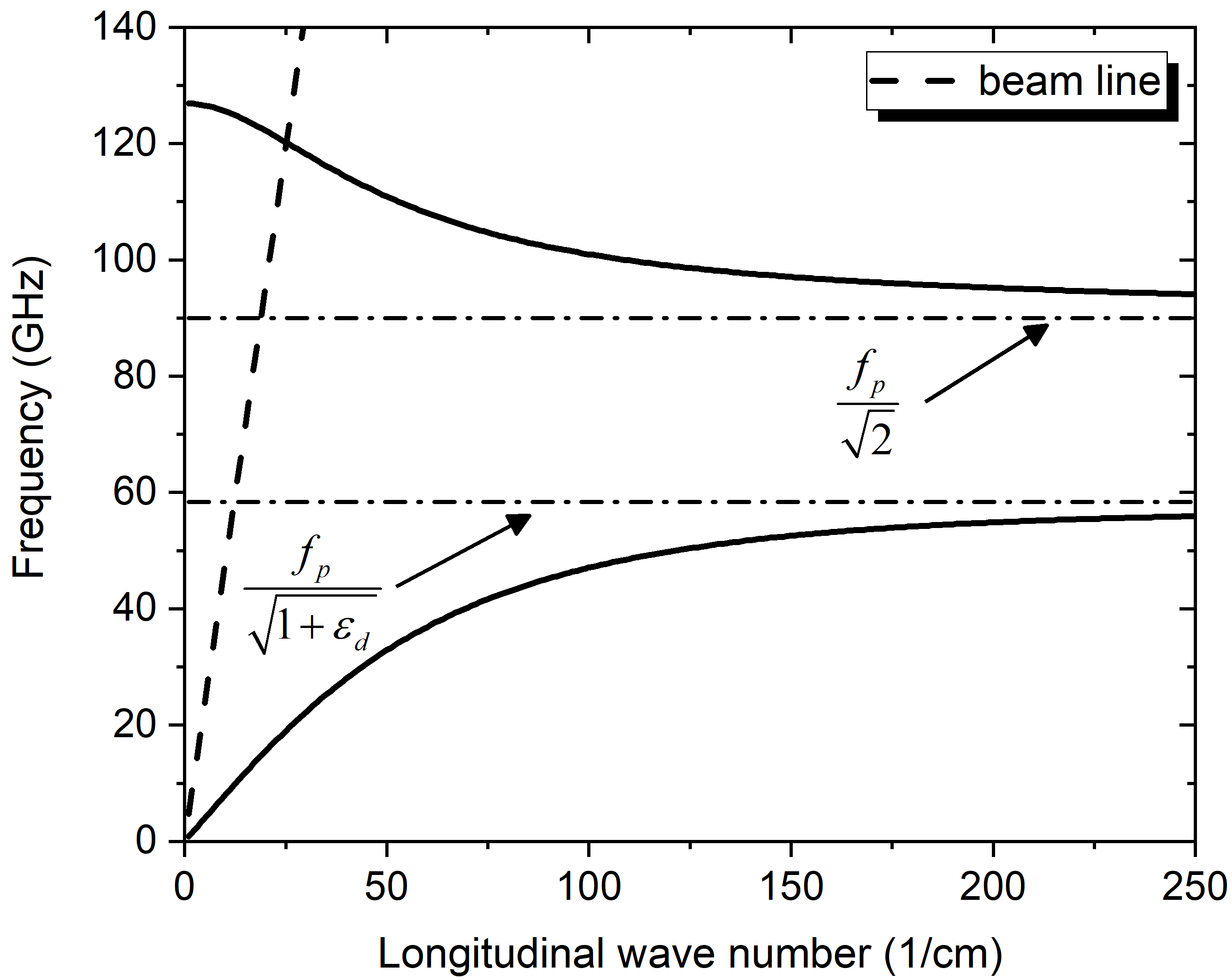}
  \caption{The dispersion of the first six TM eigenwaves of the dielectric waveguide with hollow plasma channel.}\label{Fig:02}
\end{figure}

Considering that in our case the wakefield excitation is based on the Cherenkov interaction of the relativistic drive bunch of charged particles with the eigenwaves of the waveguide under consideration, we have first of all investigated numerically its dispersion. Figure~\ref{Fig:02} shows the first six dispersion curves for the dielectric waveguide with the hollow plasma channel.
When compared to the dielectric waveguide with the vacuum channel for charged particles, the presence of the hollow plasma channel in the dielectric waveguide gives appearance to two additional eigenwaves. These are the surface waves, which can propagate along two vacuum--plasma and plasma--dielectric boundaries. The lower-frequency surface wave, having the asymptotics $f \rightarrow f_p \left/ \sqrt{1+\varepsilon_d}\right.$ at $k_z \rightarrow \infty$ , exhibits the normal dispersion law and is the forward wave. The higher-frequency surface wave having the asymptotics $f \rightarrow f_p \left/ \sqrt{2} \right.$ at $k_z \rightarrow \infty$, shows the anomalous dispersion law, and is the backward wave. The beam line with the dispersion $\omega = k_{z}v$, crosses the waveguide modes at definite frequencies. These are the resonant frequencies, which are determined by solving numerically the equation $D(\omega_s)=0$. The first four resonant frequencies are found to be 120.2 GHz, 327.6 GHz, 1040.0 GHz and 1885.5 GHz. It should be noted that for the parameters chosen for the numerical analysis, the lowest--frequency mode does not cross the beam line, and therefore, will not contribute to the excited wakefield. In the special case that the circular waveguide comprises a hollow plasma channel without any dielectric, there exist only two resonance frequencies. These are the frequencies of the surface waves that exist on the inner and outer boundaries of the plasma~\cite{Markov2008}. The presence of dielectric gives rise to additional resonant frequencies, which belong to the waves having the frequency regions with phase velocities less than the speed of light.

\begin{figure}[!th]
  \centering
  \includegraphics[width=0.49\textwidth]{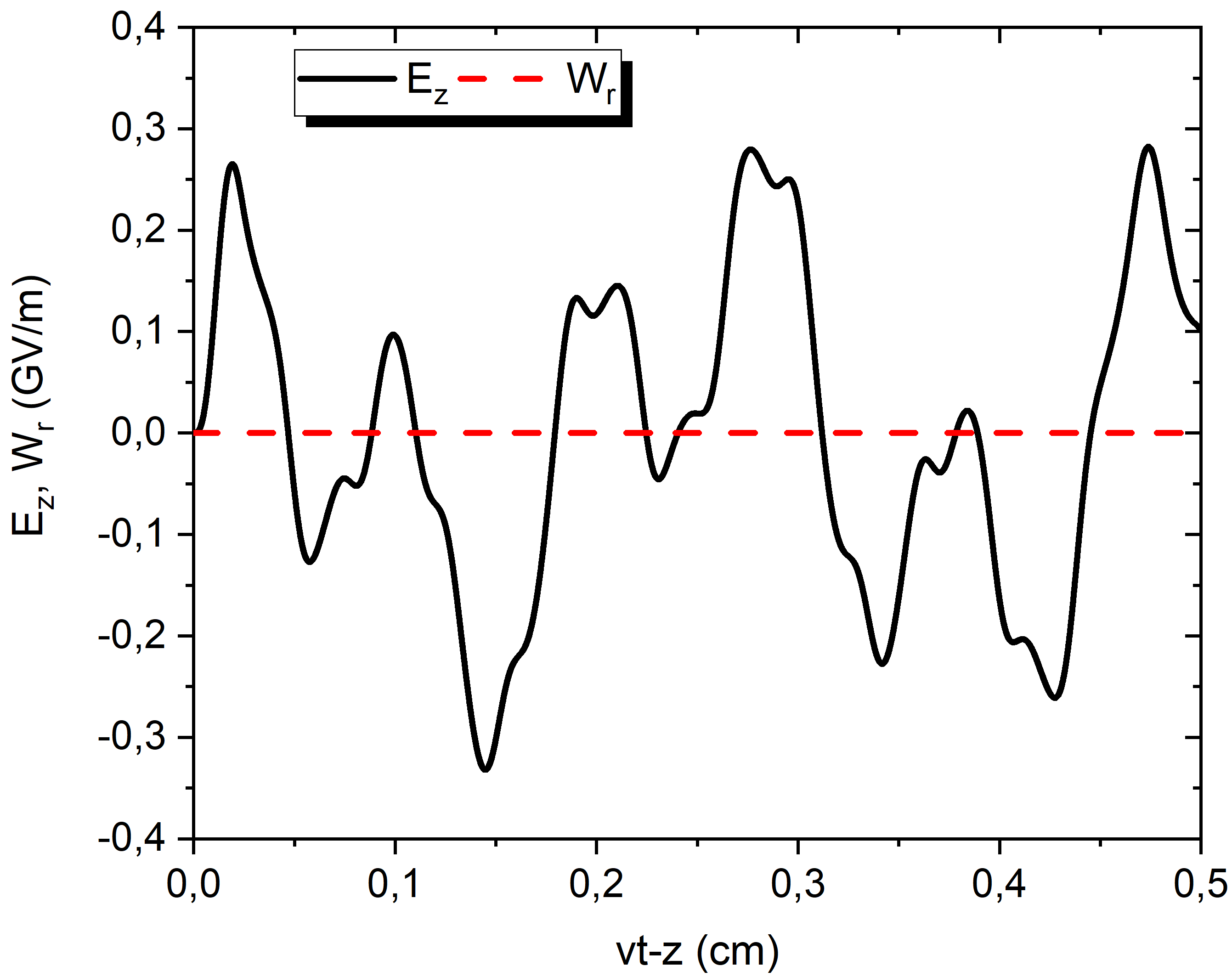}
  \includegraphics[width=0.49\textwidth]{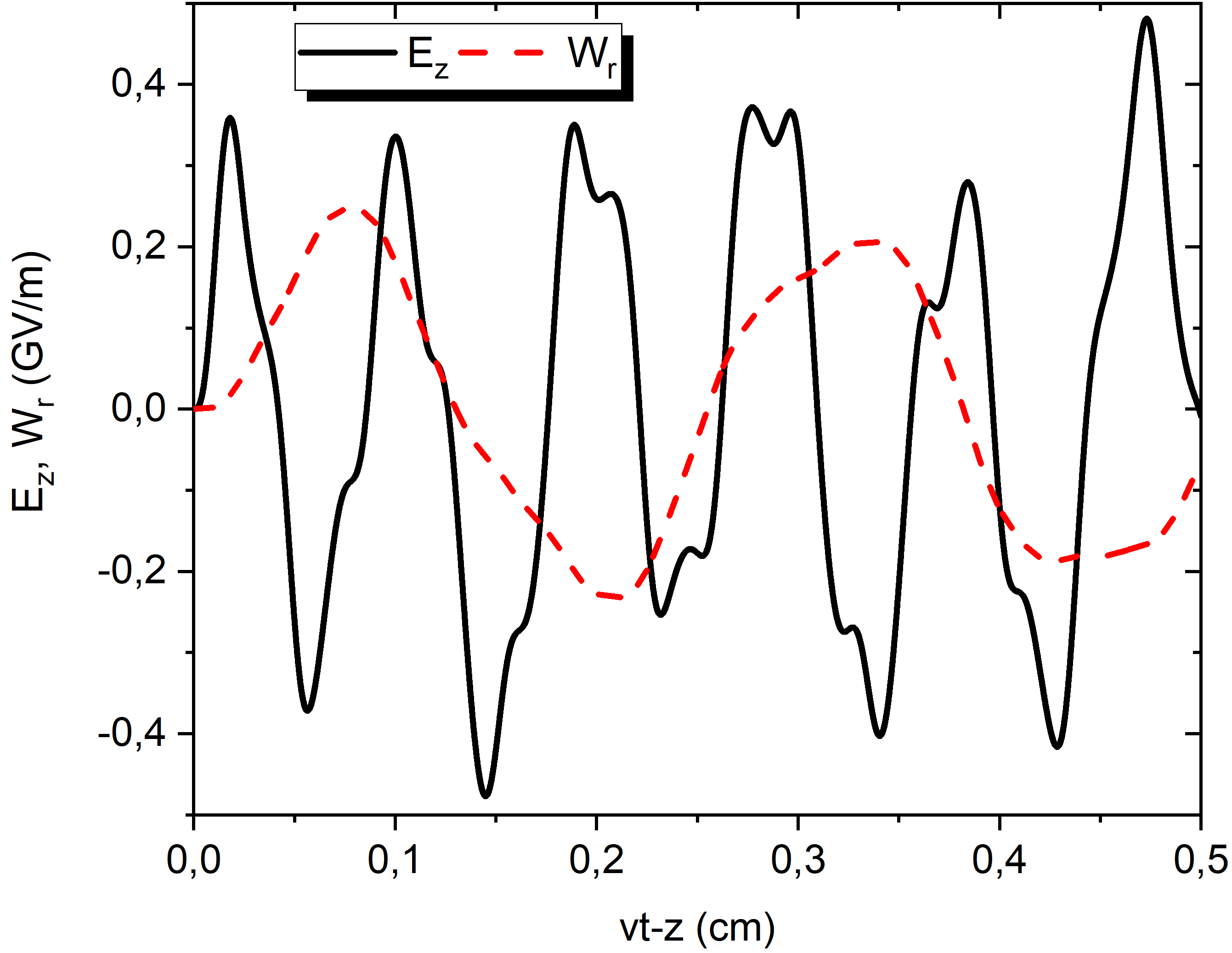}
  \caption{Longitudinal distributions of axial $E_z$ and radial $E_r-\beta H_\phi$ fields, both excited by the drive electron bunch in vacuum (left) and in plasma (right), at $r=100\,\mu m$ and at $r=350\,\mu m$ respectively. The drive bunch propagates from right to left along the waveguide axis with constant velocity. The position of the drive bunch head corresponds to $vt-z=0$.}\label{Fig:03}
\end{figure}

With knowledge of the resonant frequencies $\omega_s$ one can now turn to the numerical analysis of the excited wakefield. Figure~\ref{Fig:03} shows the longitudinal distributions of the axial wakefield $E_z$ and the radial wakefield $E_{r}-\beta H_{\phi}$ in the vacuum channel region and the plasma region, calculated at $r=100\,\mu m$ and $r=350\,\mu m$ respectively. It is obvious that the transverse wakefield amplitude distributions in the vacuum and plasma regions are essentially different, both qualitatively and quantitatively. In a vacuum channel the amplitude of the transverse wakefield is almost zero. This effect has been analytically and numerically demonstrated in~\cite{Schroeder1999,GessnerPhD}, which are dedicated to the wakefield excitation in the hollow plasma channel. The transverse wakefield amplitude has been shown to decrease with increase in the drive bunch energy, and tends to zero in vacuum region for GeV energies. In plasma, the transverse wakefield amplitude is nonzero, and exhibits nearly a monochromatic character  with weak shape distortions.
At the same time, the longitudinal wakefield amplitude is nonzero in both the plasma and vacuum regions, and has a multiwave structure in the both regions. Moreover, it is obvious that the characteristic spatial periods of longitudinal and transverse wakefields considerably differ from each other. The given structure of the drive bunch--excited wakefield components permits making some preliminary conclusions as to the transverse dynamics of the test bunch. The test bunch particles moving in vacuum will be subjected to neither focusing nor defocusing force on the side of the wakefield under acceleration. Unlike the vacuum region, in the plasma region behind the drive electron bunch, responsible for the wakefield excitation, there exist, respectively, the regions, where $E_{z}<0$, $E_r-\beta H_{\phi}>0$ (for the electron test bunch), and $E_{z}>0$, $E_r-\beta H_{\phi}<0$ (for the positron test bunch). And hence, the electron test bunch and the positron test bunch following the drive bunch can be concurrently accelerated and focused in the radial direction.

\begin{figure}[!th]
  \centering
  \includegraphics[width=0.49\textwidth]{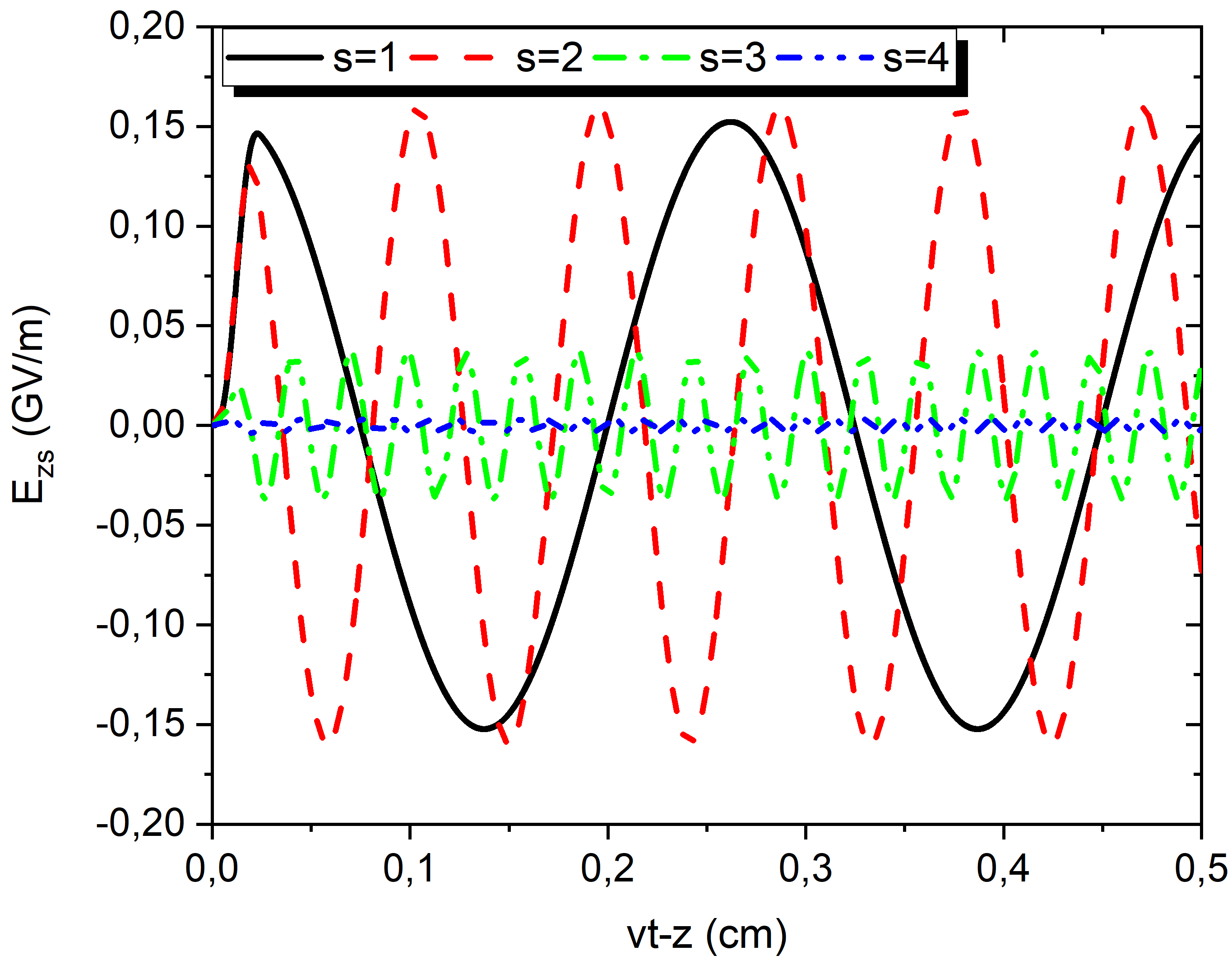}
  \includegraphics[width=0.49\textwidth]{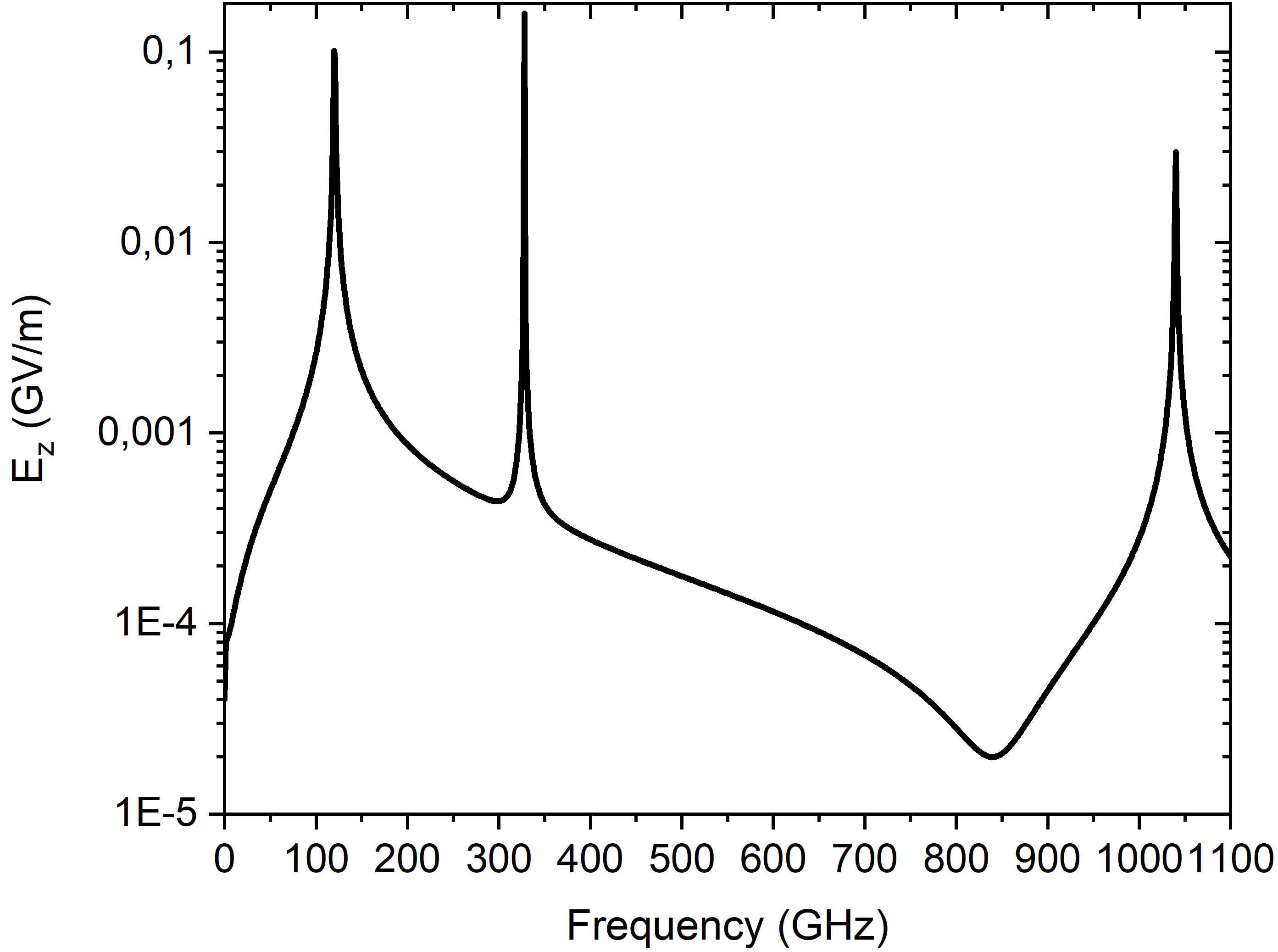}
  \caption{Longitudinal distribution of the axial wakefield of the first four eigenmodes resonant with the drive bunch $E_{z,s}$, excited in the vacuum at $r=100\,\mu m$ . The position of the drive bunch head corresponds to $vt-z=0$. The axial wakefield spectrum $E_z$, excited in the vacuum at $r=100\,\mu m$.}\label{Fig:04}
\end{figure}

\begin{figure}[!th]
  \centering
  \includegraphics[width=0.49\textwidth]{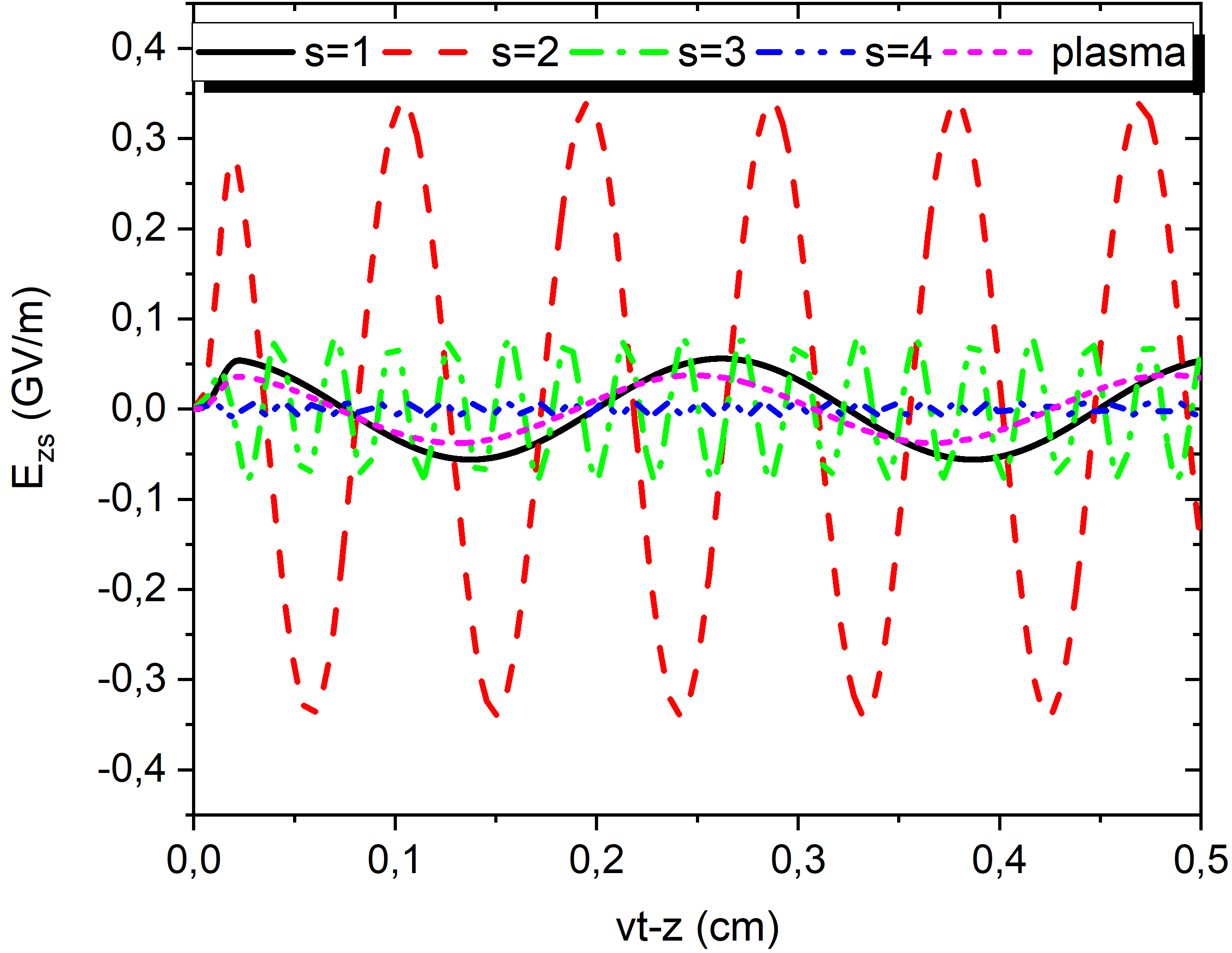}
  \includegraphics[width=0.49\textwidth]{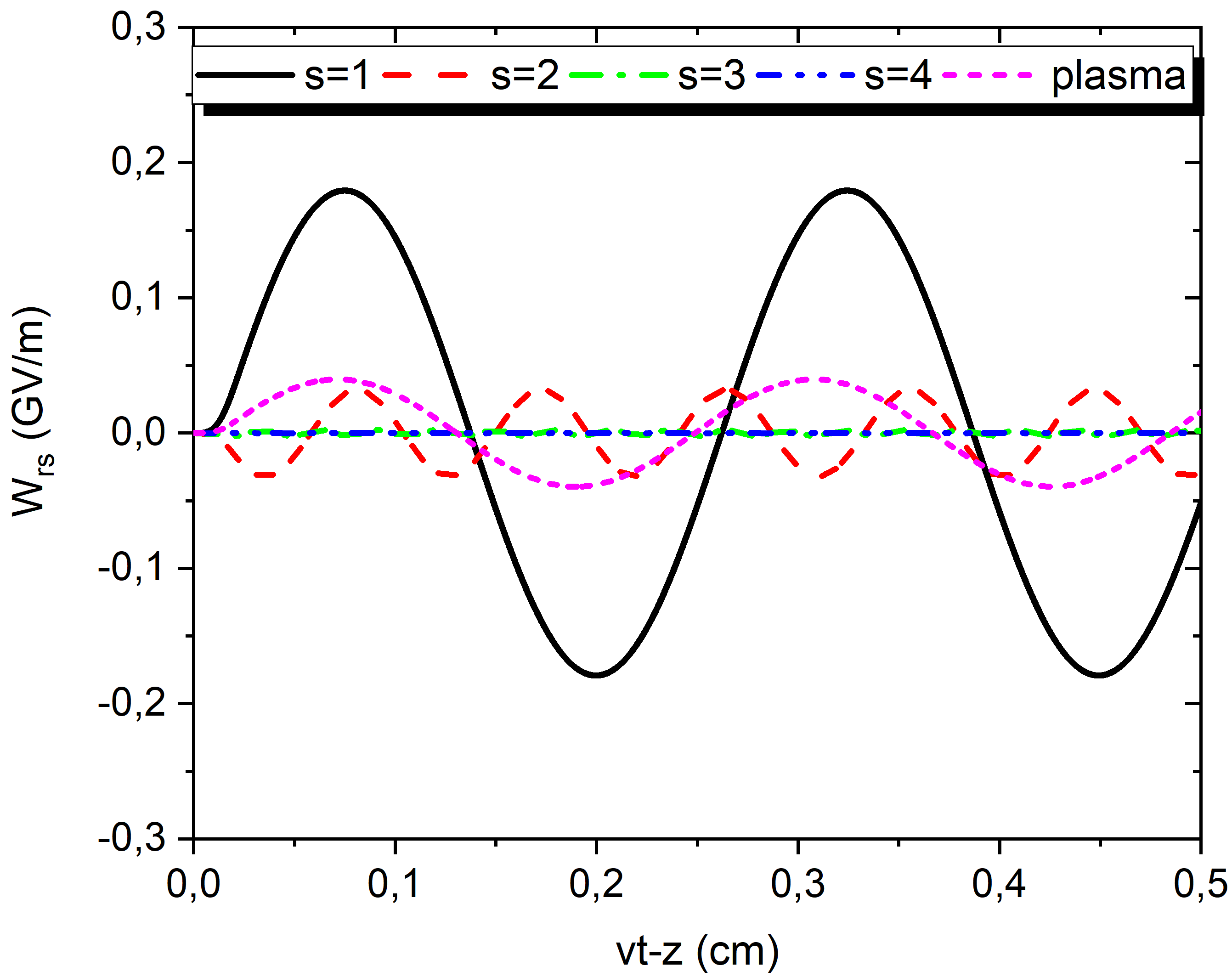}
  \caption{Longitudinal distributions of axial $E_{z,s}$ and radial $E_{r,s}-\beta H_{\phi,s}$ fields of the first four resonant modes and the ones of the plasma wave, excited in the plasma region at $r=350\,\mu m$. The position of the drive bunch head corresponds to $vt-z=0$.}\label{Fig:05}
\end{figure}

The reason for nonmonochromatic amplitudes is due to the excitation of a great number of resonant frequencies of waveguide eigenwaves. To demonstrate the effect, Fig.~\ref{Fig:04} and Fig.~\ref{Fig:05} show the longitudinal distribution of the axial $E_{z,s}$ and radial $E_{r,s}-\beta H_{\phi,s}$ fields of the first four resonant with the drive bunch modes ($s=1-4$), and also of the drive electron bunch-excited plasma wave in vacuum at $r=100\,\mu m$, and in the plasma at $r=350\,\mu m$. The mode analysis shows that the main contribution to the formation of the total wakefield comes from the first three resonant modes, and also from the Langmuir wave, which is localized in the plasma region. The contribution of higher--frequency resonant modes to the excited wakefield is negligibly small.
It is obvious that in vacuum, the amplitude of the total longitudinal field is almost equally contributed by the first two resonant modes. Their spatial periods differ from one another, and the addition results in a complex longitudinal structure of the total longitudinal field.
In the plasma region, for the longitudinal field, the amplitude is most greatly contributed by the second resonant mode ($s=2$). The amplitudes of the first ($s=1$), third ($s=3$) resonant modes, and also, the Langmuir wave amplitude are nearly the same in value, which differs from the maximum amplitude by a factor of about 4. A similar situation with the mode content is observed for the transverse field, as well. Namely, the most substantial contribution to the amplitude comes from one resonant mode, this being the first resonant mode ($s=1$). The amplitudes of the rest transverse field components (the second resonant mode ($s=2$) and the Langmuir wave) have close values, and similarly differ from the maximum amplitude value by a factor of about 4.

\begin{figure}[!th]
  \centering
  \includegraphics[width=0.49\textwidth]{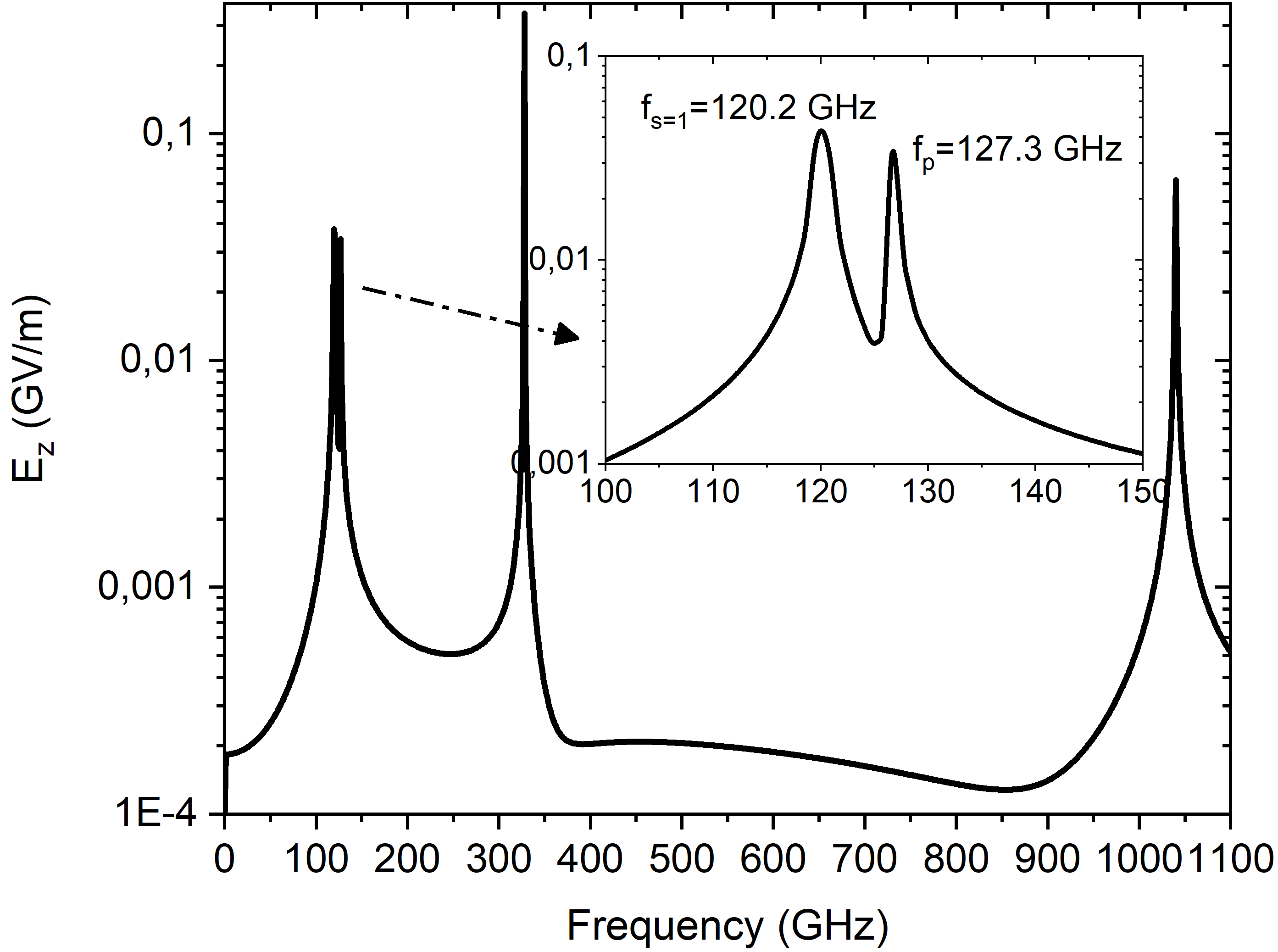}
  \includegraphics[width=0.49\textwidth]{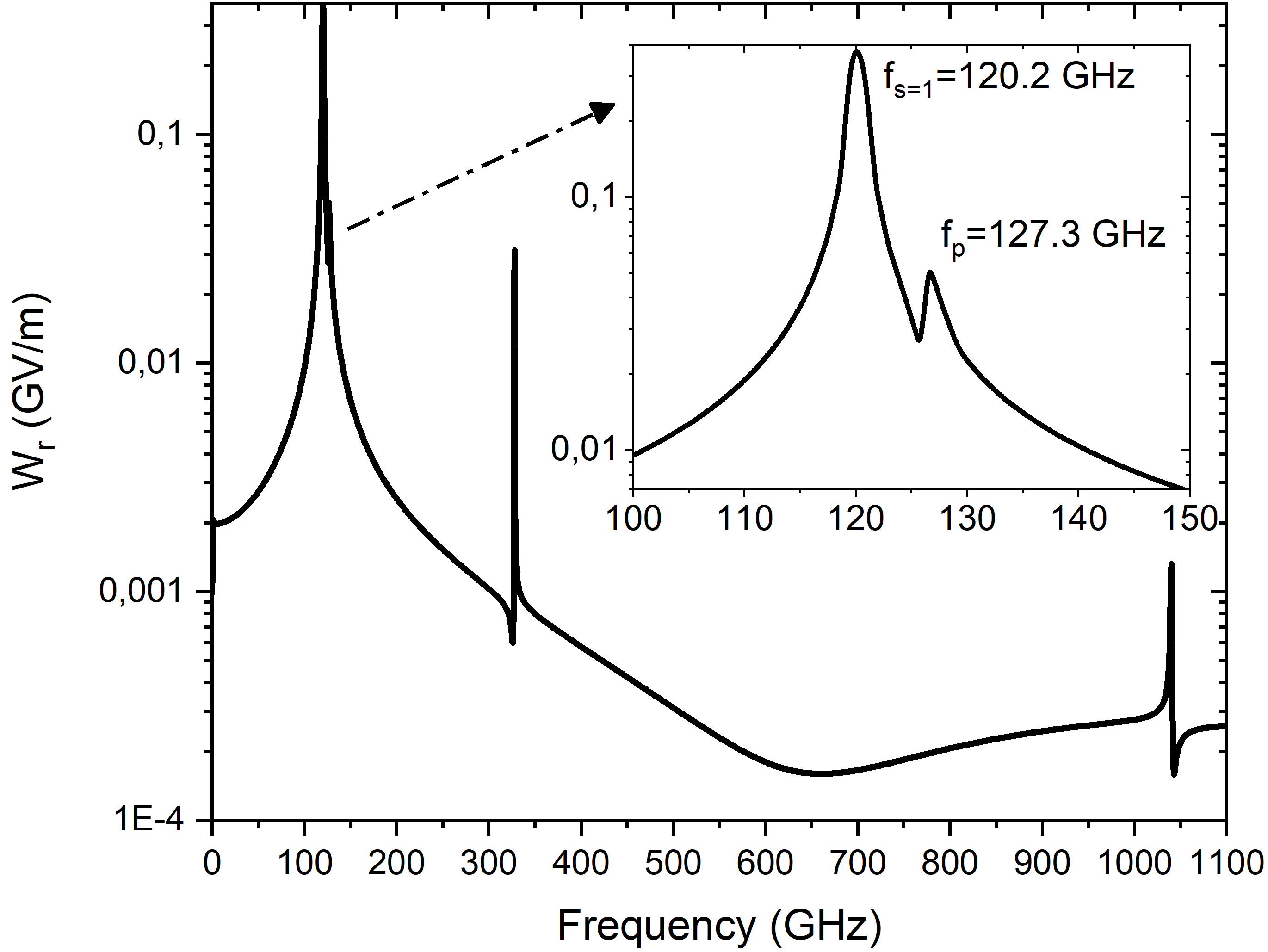}
  \caption{Spectra of axial $E_z$ and radial $E_r-\beta H_\phi$ wakefields excited in the plasma region at $r=350\,\mu m$.}\label{Fig:06}
\end{figure}

We have also performed the Fourier analysis of the spectrum for longitudinal and transverse wakefields excited in the plasma region. The results of the analysis are presented in Fig.~\ref{Fig:05} and Fig.~\ref{Fig:06}, which show the corresponding spectra.
The frequencies, at which the spectral amplitudes are nonzero, represent the resonant mode frequencies, and also, the Langmuir wave frequency. The Fourier analysis data are in complete agreement with the results of the mode analysis, and taken together, they give us an insight into the internal structure of the excited wakefield as far as the amplitude-frequency content is concerned. Besides, the numerical analysis has demonstrated that a further increase in the number of resonant modes considered causes no changes in the profiles of either longitudinal or transverse wakefields. In turn, this permits us to estimate the maximal frequency value, which should be taken into account in the computational codes, where to find the excited electromagnetic field, use is made of the Fourier expansion (for example CST Studio). It should be noted that in previous numerical studies based on the PIC--simulation~\cite{Markov2022}, the spectrum analysis was not performed. According to the results of those studies, it was concluded that the excited transverse wakefield (and hence, the focusing force for the test bunch) was mainly contributed by the plasma Langmuir wave at the plasma frequency $\omega_p$. The spectrum analysis made on the basis of the developed linear theory has demonstrated that the main contribution to the transverse wakefield comes not from the plasma Langmuir wave but from the backward plasma surface wave with its resonant frequency close to the frequency of the plasma Langmuir wave (for the given parameters of both the plasma--dielectric waveguide and the drive bunch).

The derived expressions make it possible to obtain analytical estimation of the longitudinal and transverse wakefields, and that, in turn, permits estimation the energy gain of the test bunch, the drive bunch energy losses, and also, the focusing of the both bunches (in the approximation of neglecting the inverse effect of the excited wakefield on bunch particles). Using the obtained Green's functions we can derive the expressions for the wakefields  excited by solid and annular drive bunches that propagate either through the vacuum channel, or through the plasma, or, in the general case, by overlapping completely with the cross--section of the charged particle transport channel.

\begin{figure}[!th]
  \centering
  \includegraphics[width=0.49\textwidth]{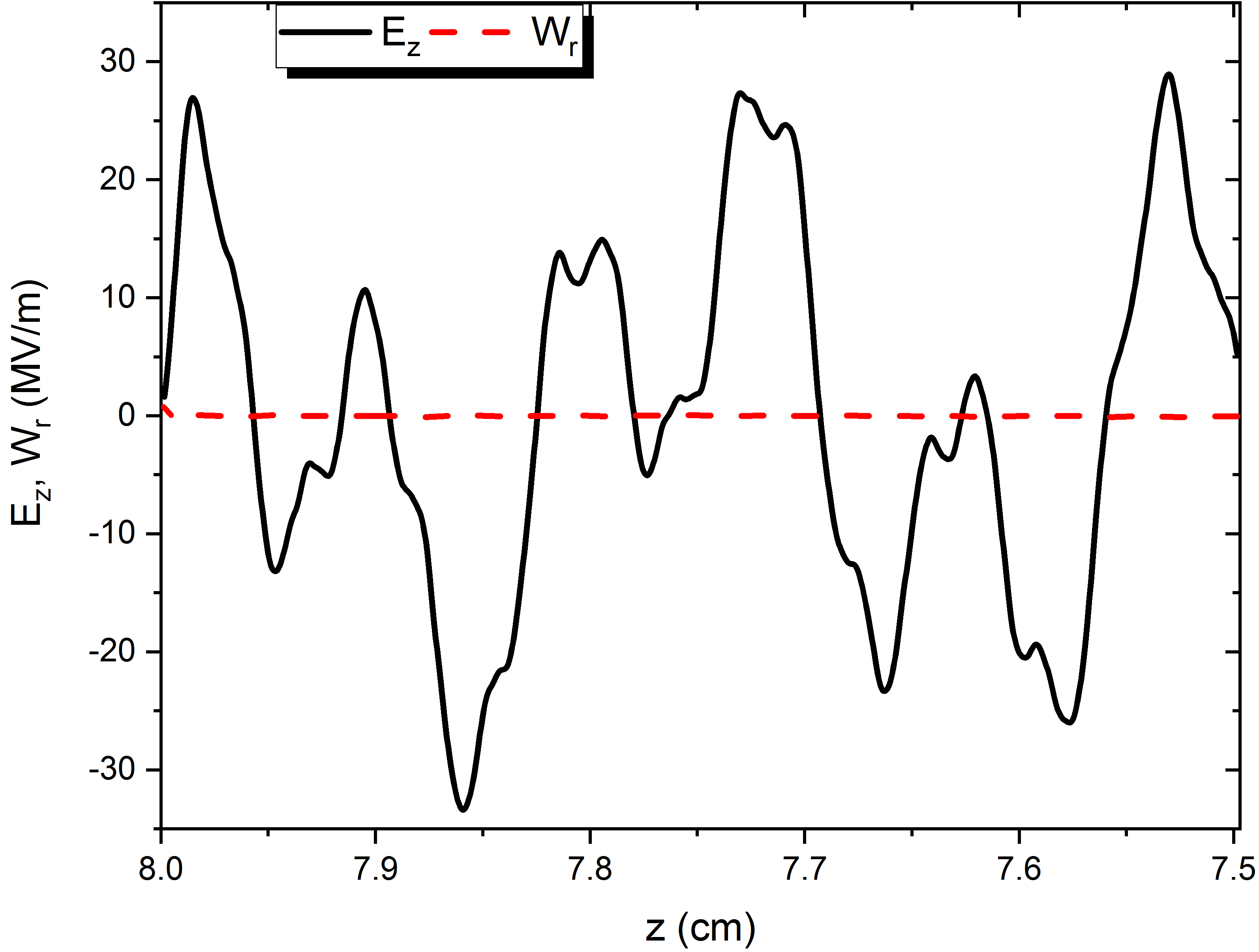}
  \includegraphics[width=0.49\textwidth]{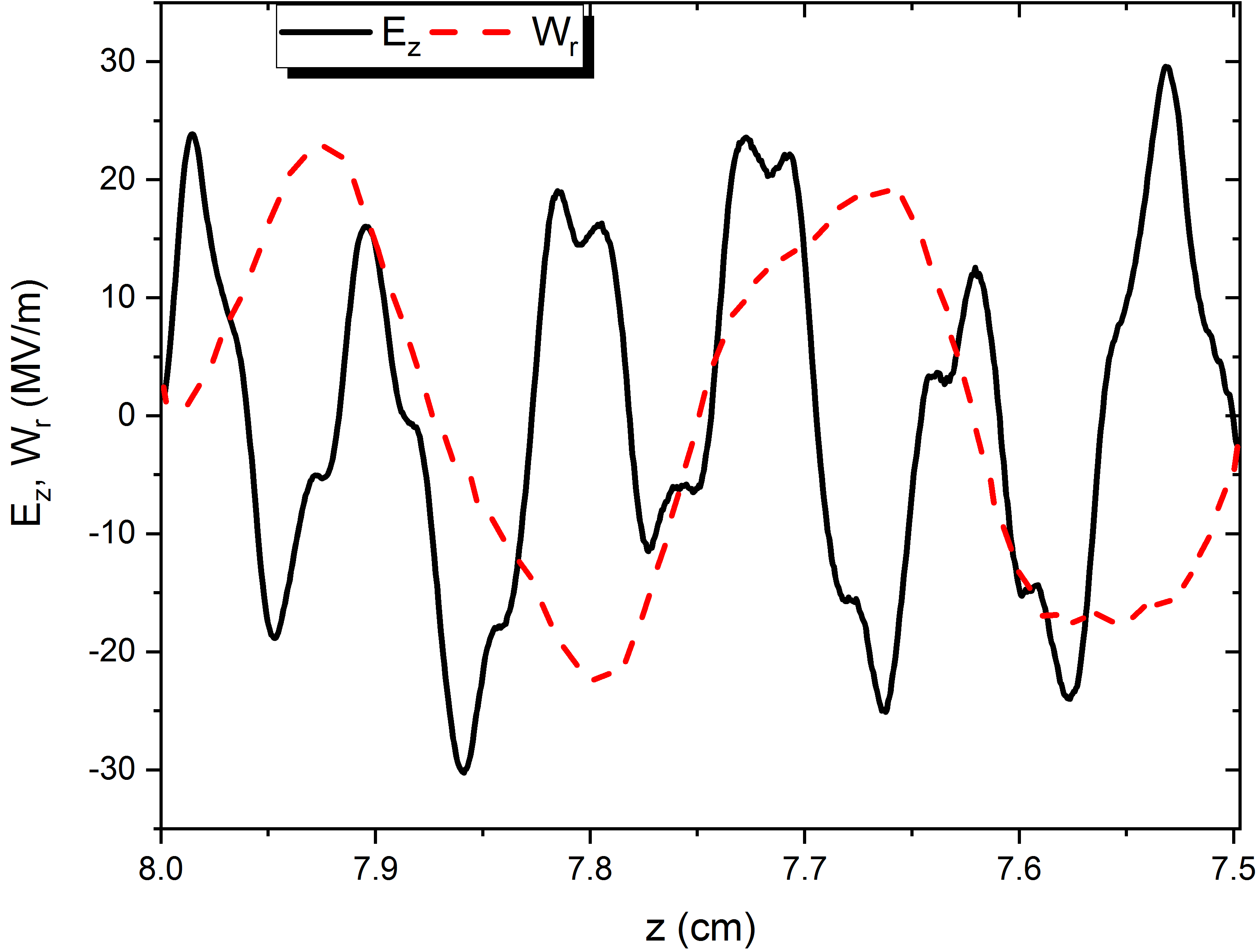}
  \caption{Resulting--from--simulation longitudinal profiles of the electron bunch--excited axial $E_z$ and radial $E_r-\beta H_\phi$ wakefields in vacuum (left) and plasma (right) regions at $r=100\,\mu m$ and $r=350\,\mu m$ respectively. The drive bunch travels from right to left along the waveguide axis with constant velocity. The position of the drive bunch head corresponds to the longitudinal coordinate $z=8$.}\label{Fig:07}
\end{figure}

\begin{figure}[!th]
  \centering
  \includegraphics[width=0.49\textwidth]{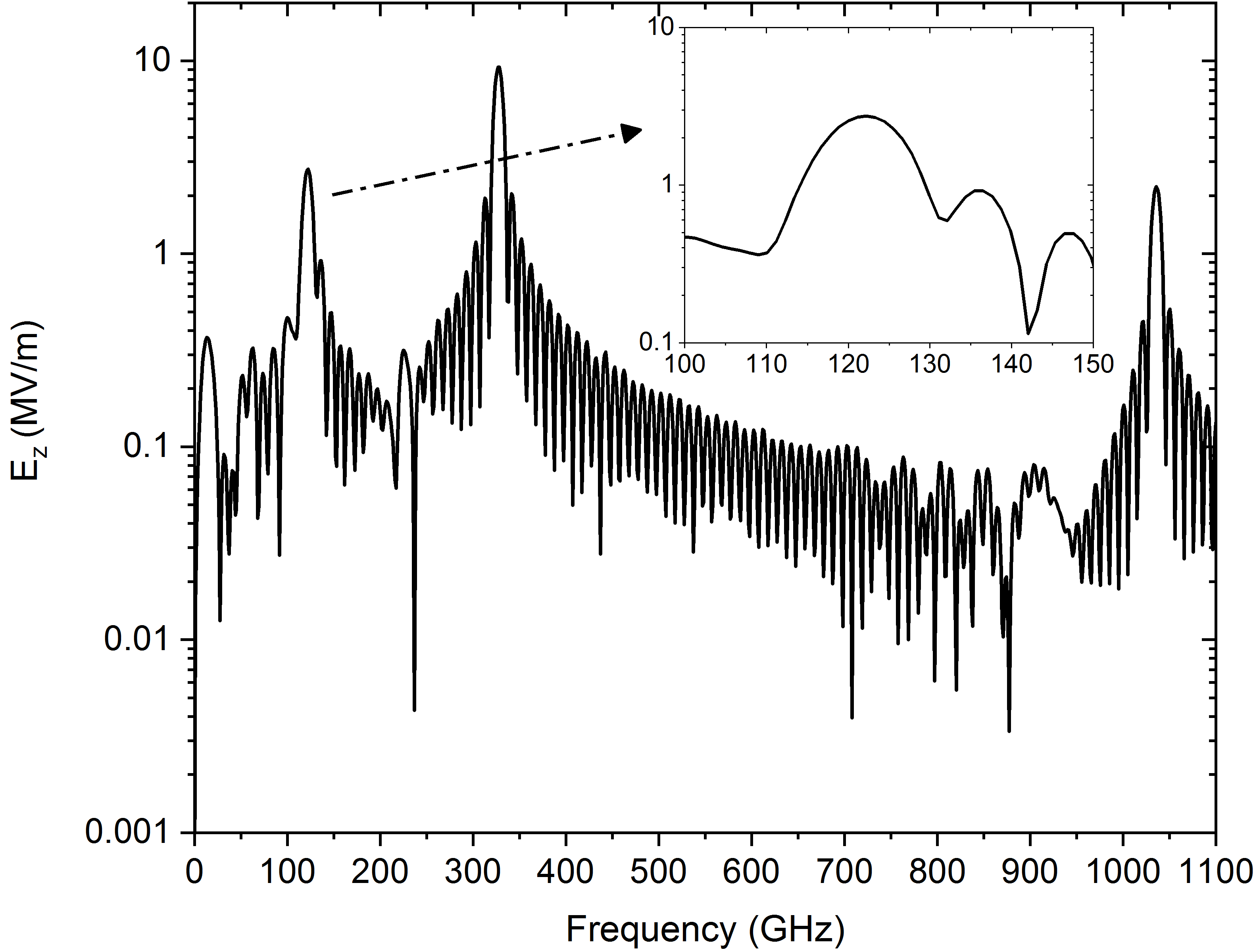}
  \includegraphics[width=0.49\textwidth]{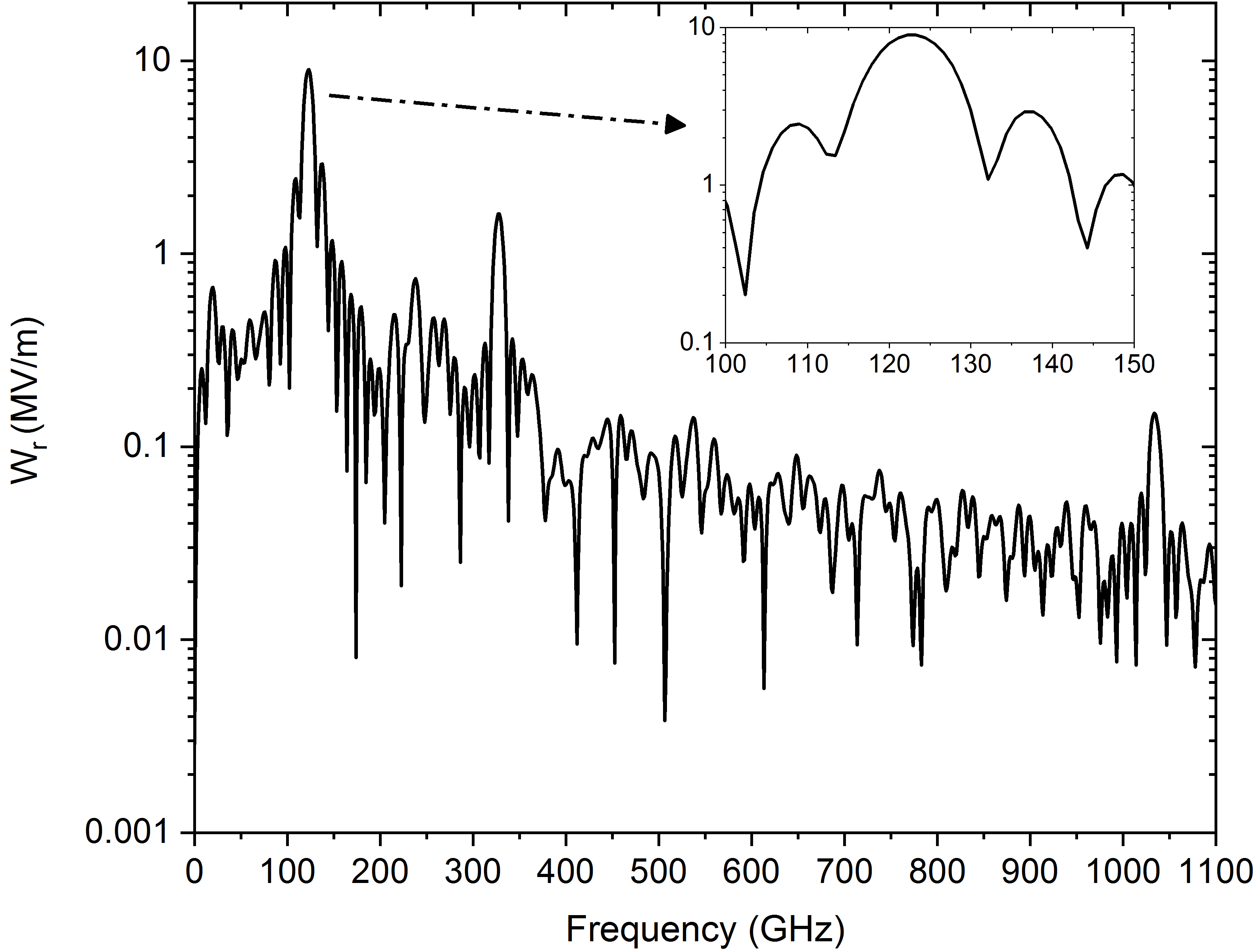}
  \caption{Simulation spectra for the axial $E_z$ and radial $E_r-\beta H_\phi$ wakefields excited in the plasma region at $r=350\,\mu m$.}\label{Fig:08}
\end{figure}

The data obtained on a basis of the presented linear theory were compared with the data of simulation of quasi--linear regime ($n_{b}/n_{p}=0.075$) of the process of wakefield excitation by the drive electron bunch. To get this regime the charge of the drive bunch has been reduced by ten time. For simulation, we have used the native PIC--based code, which was earlier verified with both the widely used code XOOPIC~\cite{VERBONCOEUR1995199}, and the data of the previous analytical studies~\cite{Sotnikov2014,GALAYDYCH2022}. Figure~\ref{Fig:07} demonstrates the resulting from simulation longitudinal distributions of  the axial field $E_z$ and the radial field $E_{r}-\beta H_{\phi}$ in the vacuum channel region and in the plasma region, calculated at $r=100\,\mu m$, and $r=350\,\mu m$ respectively.
The analytical theory data (Fig.~\ref{Fig:03}) and the simulation results (Fig.~\ref{Fig:07}) show good qualitative agreement between longitudinal axial and radial wakefield profiles in both the vacuum and plasma regions. It should be noted that this agreement is observed for the $vt-z$ values from the drive bunch tail position up to the position of the group front of the excited wakefield~\cite{Balakirev2001,Balakirev2003}. For $vt-z$ values from the position of the group front of wakefield  up to the position of the accelerating structure input, the profiles under consideration differ significantly. Among the main reasons for the difference between the results gained with the developed linear theory and the ones obtained by simulation we may mention the following: (I) the linear theory assumes the absence of plasma density perturbation and, consequently, the plasma electrons falling into the vacuum channel. That takes place (though slightly) in simulation even for the linear regime; (II) the presented theory has been constructed in the approximation of the longitudinal unboundenness of the accelerating structure, and this, in turn, may cause the difference in the frequency spectrum of the excited wakefield in comparison with the field spectrum estimated in the simulation. The results of the spectrum analysis for the longitudinal and transverse wakefields excited in the plasma region at $r=350\,\mu m$ and estimated by the numerical simulation are presented in Fig.~\ref{Fig:08}.
A comparison between the spectrum analysis data obtained from the analytical theory (Fig.~\ref{Fig:06}) and from simulation (Fig.~\ref{Fig:08}) shows that there is some difference in the frequency content of the excited wakefields. The simulation displays a more multifrequency spectrum , but the frequencies, to which the major maxima of the spectrum correspond, are almost the same. A great many nonresonant frequencies can be treated as a transient radiation field, which arises at the drive bunch injection into the accelerating structure, and which is not taken into account for the longitudinally unbounded waveguide statement of the problem. And yet, considering that the numerically simulated longitudinal amplitude profiles of  the axial and transvers wakefields, where these frequencies get excited and are present in the corresponding spectra, are not too different in quality from the corresponding amplitude profiles derived from the analytical theory, it can be concluded that the contribution of these frequencies to the total wakefield is not essential (for the $vt-z$ values from the position of the drive bunch tail to the group front position of the excited wakefield). Surely, these frequencies do contribute to the total amplitude of the wakefield components, but they do not essentially affect the longitudinal profile of these amplitudes.

The PIC--based numerical calculation of the drive bunch--excited wakefields distribution in the plasma--dielectric wakefield structure is a very time-consuming procedure. To provide good resolution in the excited oscillation spectra, one must simulate lengthy systems with a great number of macroparticles. For this reason, the above-given analytical results can be used for a rapid choice of the parameters to perform further more exact numerical computations.

\section{Conclusions}\label{section:5}
In paper we have developed a linear theory of wakefield excitation by a drive charged particle bunch in a cylindrical dielectric waveguide with  a hollow plasma channel. Green's functions have been built for the cases, where the point--like particle exciting the wakefield travels in both the vacuum channel and the plasma region. Based on the constructed theory, we have performed numerical studies on the amplitude structure of wakefield components excited by the finite--size drive bunch. The studies have demonstrated the possibility of radially-stable acceleration of test bunches of both types (electron and positron). The mode and spectrum analyses of the excited wakefield have shown that to obtain analytical estimations for the field amplitudes and their space--time structure, it will suffice to take into account several first resonant eigenmodes together with the plasma Langmuir wave (in the plasma region). For the linear regime,  expressions for the longitudinal and transverse wakefields were earlier derived for two transversely unlimited cases. One case is in~\cite{Schroeder1999}, where the radius of the vacuum region is finite, while the plasma thickness is infinite; the second case is in~\cite{Gessner2016} for the hollow plasma channel of finite thickness surrounded by vacuum, which extends to infinity. Unlike the mentioned cases, the present study is concerned with the accelerating structure of finite transverse size, having an additional dielectric layer. It has been demonstrated that in the vacuum region the transverse wakefield is almost absent (for high bunch energies), whereas in the plasma region the transverse wakefield is nonzero and can serve as focusing for the electron and positron test bunches. The simulation data are in good agreement with the data of the developed  linear analytical theory.

\section*{Acknowledgments}
The study is supported by the National Research Foundation of Ukraine under the program “Leading and Young Scientists Research Support” (project \# 2020.02/0299).

\bibliographystyle{elsarticle-num}
%\bibliography{Bibliography_NIMA}	
%\bibliography{Bibliography}
\bibliography{BibliographyURL}
\end{document}